\pdfoutput=1
\documentclass[epj]{svjour}
\usepackage{amsmath,amssymb}
\usepackage{graphicx}
\newcommand{\fig}[2]{\centerline{\includegraphics[width=#1\columnwidth]{./figures/#2}}}
\newcommand{\Fig}[1]{\centerline{\includegraphics[width=\columnwidth]{./figures/#1}}}
\newcommand{\nn}{\nonumber}
\newcommand{\rmd}{\mathrm{d}}
\newcommand{\rme}{{e}}
\newcommand{\rGamma}{{\mathrm{\Gamma}}}

\begin{document}
\title{Elasticity of a contact-line and avalanche-size distribution at depinning}

\author{Pierre Le Doussal  
\and Kay J\"org Wiese 
}                     
%
%
\institute{CNRS-Laboratoire de Physique Th\'eorique de l'Ecole Normale
Sup\'erieure, 73231 Paris  Cedex 05, France 
}
\date{August 27, 2009 }
%
\abstract{Motivated by recent experiments, we extend the Joanny-deGennes calculation of the elasticity of a contact line to an arbitrary contact angle and an arbitrary plate inclination in presence of gravity.
This requires a diagonalization of the elastic modes around the non-linear equilibrium profile, which is carried out exactly. We then make detailed predictions for the avalanche-size distribution at quasi-static depinning: we study how the 
universal (i.e.\ short-scale independent) rescaled size distribution and the ratio of moments of local to global avalanches depend on the precise form of the elastic kernel.
} 
\maketitle
\section{Introduction}
\label{intro}

\begin{figure}
\fig{1}{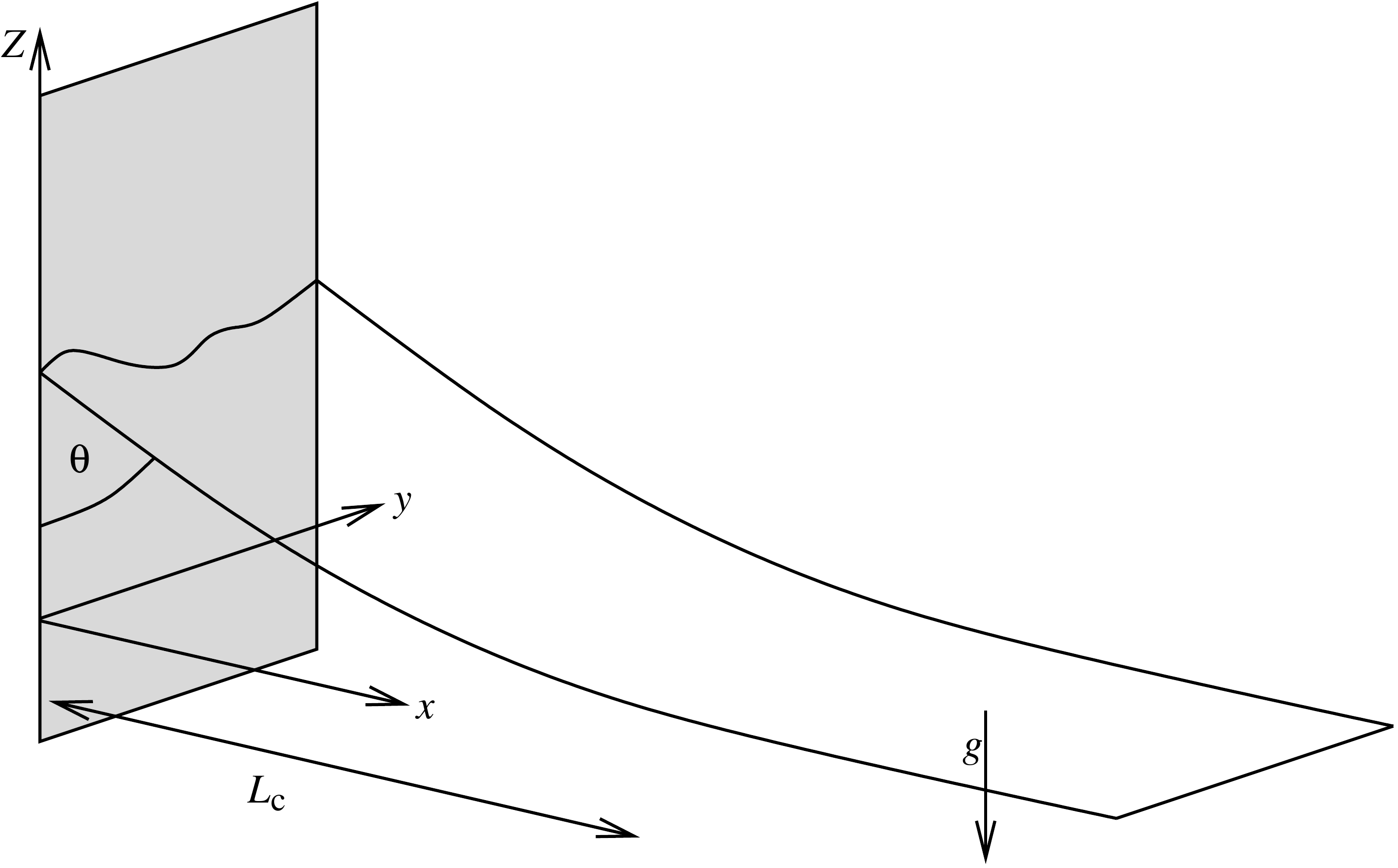}
\caption{The coordinate system for a vertical wall. The air/liquid interface becomes flat for $x\gg L_c$.}
\label{a11}       
\end{figure}

\begin{figure}
\fig{0.7}{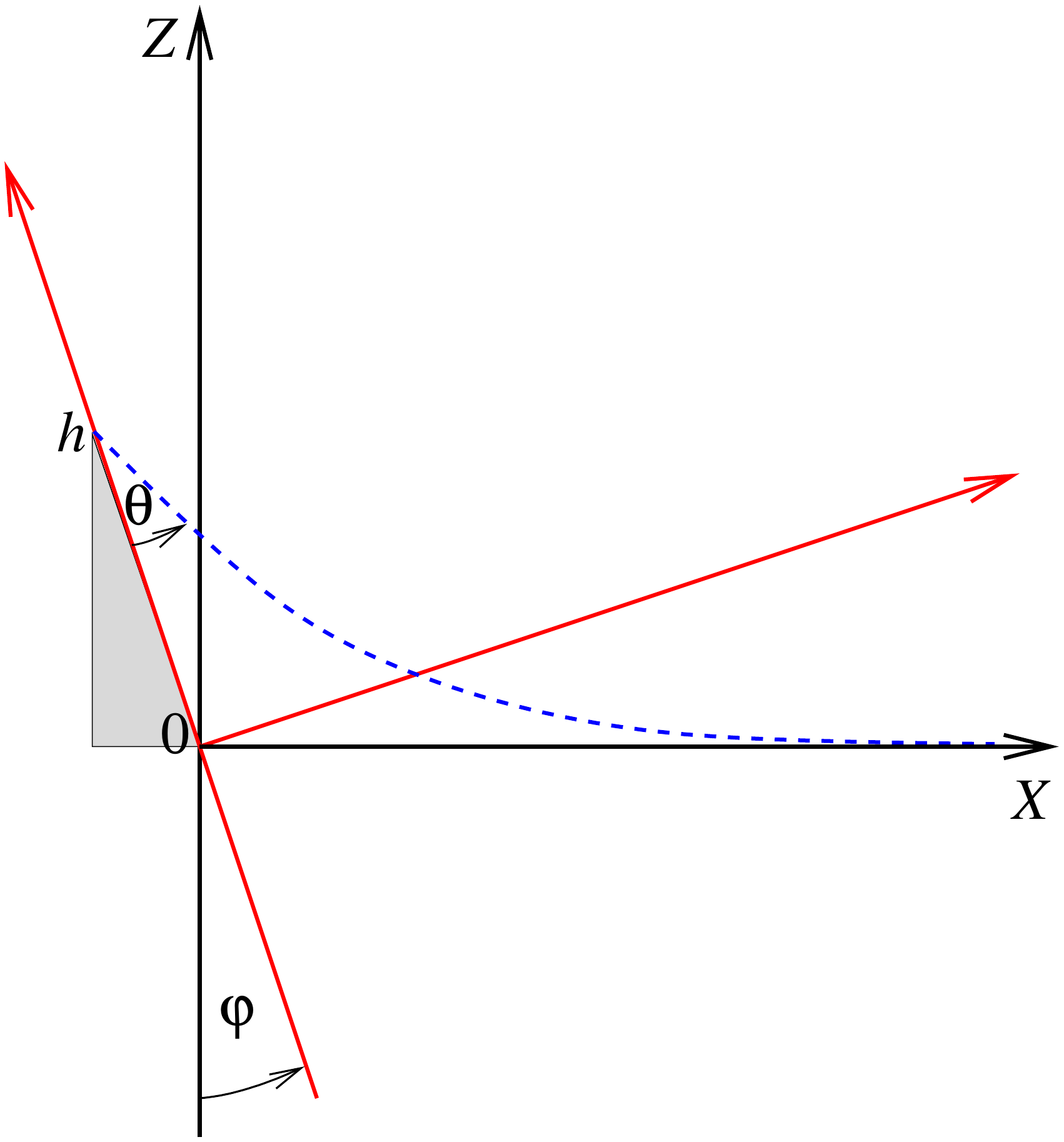}
\caption{The coordinate system for a wall inclined by $\varphi$ with respect to the
vertical: the coordinate $y$, not shown, is perpendicular to the plane of the figure. The gray shaded volume is not filled by the liquid,  leading to the gravity subtraction (second term in second-to-last line of eq.~(\ref{a5e}).)}
\label{figk1}
\end{figure}

Recent experiments on contact-line dynamics near depinning \cite{LeDoussalWieseMoulinetRolley2009} have allowed for a direct measurement of the renormalized disorder correlator, and of the avalanche-size statistics. It was concluded in \cite{LeDoussalWieseMoulinetRolley2009} that the precise form of the contact-line elasticity may have an important impact on the distribution of avalanche sizes. Thus, a detailed
prediction for the former is needed. Surprisingly, a review of the literature reveals that this problem appears to have been treated only neglecting gravity 
\cite{JoannyDeGennes1984}\footnote{
\cite{JoannyDeGennes1984} introduce the capillary length $L_c$ ad hoc as a cutoff for the contact-line profile in response to a $\delta$-like force.}, or in presence of gravity only in the case where the equilibrium configuration of the interface is flat and horizontal, as for a vertical wall and a contact angle $\theta$ of $90^\circ$
\cite{PomeauVannimenus1985,NikolayevBeysens2003}, see Figs. \ref{a11} and \ref{figk1}. In the latter case, the elastic energy as a function of the Fourier-transformed height profile $h_q$ takes the form:
 \begin{equation}\label{Esimple}
{\cal E}[h]= \frac12 \int \frac{dq}{2 \pi} ~ \epsilon_q \, |h_{q}|^2 \ , \qquad \epsilon _q= \gamma \sqrt{q^2+\kappa^2}\ , 
\end{equation}
with $\kappa$ the inverse capillary length, and up to terms of order $h^3$. In other cases, interpolation formulas have been proposed \cite{BrochardGennes1991}.  

There is numerical evidence \cite{Checco2009} that the contact-line energy (\ref{Esimple}) does not correctly describe the shape of the contact line depinning from a single defect. Surprisingly, the form of $\epsilon_q$ as given by (\ref{Esimple}) seems not to have been questioned, rather the discrepancies were attributed to higher-order terms in $h$. The latter are indeed present \cite{GolestanianRaphael2001,LeDoussalWieseRaphaelGolestanian2004,BachasLeDoussalWiese2006}, even for $\theta=90^\circ$ and a vertical wall ($\varphi=0$), but for small perturbations they are subdominant w.r.t.\ the dominant term (\ref{Esimple}). 

In the first part of the present article (section \ref{section3}) we show that for contact-angles leading to a non-flat profile as plotted in figure \ref{a11}, or for an inclined wall in figure \ref{figk1}, the elastic energy (\ref{Esimple}) cannot be used, but must be replaced by the more general form \begin{eqnarray}\label{final1}
&&  \frac{\epsilon_{q}}{\kappa \gamma} = \frac{ \sin (\theta ) \cos (\varphi )}{
 t} + \frac{\left(r^2-1\right)\left [t
 (r+t)+1\right]\sin ^2(\theta )}{ t \left(r^2+3 r t+3 t^2-1\right)} 
  \qquad \\
 &&  t= \sqrt{\frac{\sin(\theta +\varphi )+1}{2}} \quad , \quad  r= \sqrt{1+\frac{q^{2}}{\kappa^{2}}} 
\ . \nonumber
\end{eqnarray}
Then we discuss how this form can be measured from the contact line profile in presence
of a single defect. This section is pedagogical and self-contained and can be read with no prior knowledge in either wetting or disordered systems. 

In the second part of the paper (section \ref {section4}) we examine the consequences of the form (\ref{final1}) of the elastic kernel $\epsilon_q$ for the avalanche statistics of a con\-tact-line at depinning, i.e.\ a con\-tact-line advancing on a disordered substrate, when the fluid level is adiabatically increased. We calculate the local (i.e.\ at a given point) as well as global avalanche-size distribution, including the scaling functions. We work at quasi-static depinning using an over-damped equation of motion. The point of the present work is to study the effects arising from the precise form of the elastic kernel. We therefore assume that details of the dynamics as well as non-linearities can be neglected in comparison. We use the methods introduced in \cite{LeDoussalWiese2008c} to compute the avalanche-size distribution from the Functional Renormalization Group (FRG) theory of pinned elastic systems (see also Refs.\ \cite{WieseLeDoussal2006} for a pedagogical introduction  and \cite{LeDoussalMiddletonWiese2008,RossoLeDoussalWiese2009a} for early numerical tests of this theory). We recall and analyze the formula given in these references, hence this section can be read with no prior knowledge of disordered systems; however understanding its foundations requires the knowledge of the above mentioned literature. Then we derive a general formula for the size distribution for an arbitrary elastic kernel, not given priorly, and compute universal ratios of local-avalanche-size moments versus global ones. Finally,  we apply our general result to  the contact line with the elasticity (\ref{final1}).

\section{Model}
\label{model}

Consider a fluid in a semi-infinite reservoir, bounded by a planar plate, in presence of gravity. 
The plate is inclined by an angle $- \pi/2 < \varphi < \pi/2 $ with respect to the vertical, see figure \ref{figk1}. We consider the coordinate system $X,y,Z$ where 
$Z$ is along the vertical direction, and the equation of the plate is $Z=-X \cot \varphi$. The contact line of the fluid
is parameterized by $(h(y),y)$ along the plate, hence it is at $(X=- h(y) \sin \varphi, y, Z=h(y) \cos \varphi)$ in our coordinate system. The fluid occupies the space $Z < Z(X,y)$ and $X > -Z \tan \varphi$, hence also $X> - h(y) \sin \varphi$, where $Z(X,y)$ is the height of the  
fluid-air surface. We choose the coordinate $Z$ so that the reservoir level is $Z=0$ hence $Z(X \to \infty,y)=0$. 

The energy $E$ is the sum of the fluid-air interface energy, proportional to the area, and the gravitational energy. 
It is a functional of $Z(X,y)$, with explicit dependence on $h(y)$. Its full expression reads:
\begin{eqnarray}\label{a5e}
\lefteqn{{\cal E}[Z,h] := \frac{E[Z,h]}{\gamma}}\nonumber\\
&& = \int \rmd y \!\int\limits_{X>-h(y) \sin \varphi} \!\!\!\!\!\!\rmd X \bigg[ \sqrt{1+
[\partial_{X} Z]^{2}+[\partial_{y} Z]^{2} } +\frac{1}{2} \kappa^2 
Z^{2}\bigg]  \nonumber \\
&&  \qquad -  \frac{1}{2} \kappa^2 \int \rmd y \int^{0}_{ - h(y) \sin \varphi } \rmd X \,  X^2\, \cot^2 \varphi \nonumber
\\
&&\qquad  - \int \rmd y \int_{X>0} \rmd X \, \, 1 \nonumber\\
&&\qquad - \cos \theta \int \rmd y\,  h(y)\ .
\end{eqnarray}
$\gamma$ is the surface tension, and
\begin{equation}
\kappa = \frac{1}{L_c} = \sqrt{\frac{\rho g}{\gamma}}
\end{equation}
defines the capillary length $L_c$. We call ${\cal E}$ the reduced energy. The first term is the area of the fluid surface; the second is
the gravitational cost in potential energy for bringing a fluid element from infinity at level $Z=0$, filling the reservoir up to height $Z(X,y)$. The second line takes into account that for $\varphi > 0$ and $h(y)>0$ (resp. $\varphi<0$ and $h(y)<0$) the volume element $0<Z<-X \cot \varphi$ (resp. $-X \cot \varphi<Z<0$) is actually not filled by the liquid; this is the gray shaded region on figure \ref{figk1}. Similarly for $\varphi > 0$ and $h(y)<0$ (resp. $\varphi<0$ and $h(y)>0$) there is a corresponding missing volume element which must be added. The third line is the subtraction of the surface energy of the flat profile $h(y)=0$ (i.e.\ $Z=0$ and $X>0$), necessary to make the problem well-defined. 
The last term comes from the difference between solid-fluid $\gamma_{\mathrm{SF}}$ and solid-air $\gamma_{\mathrm {SA}}$ surface energies which defines the {\it equilibrium} contact angle, denoted $\theta$, via the usual relation $\gamma \cos \theta = \gamma_{\mathrm {SA}}-\gamma_{\mathrm {SF}}$. The definition of the contact line implies the additional boundary condition
\begin{equation} \label{constraint}
Z\left(- h(y) \sin \varphi, y\right) = h(y) \cos \varphi\ .
\end{equation}
The energy of a given configuration $h(y)$ of the contact line is obtained as $E[h]:=E[Z_h,h]$ where $Z_h(X,y)$ minimizes ${\cal E}[Z,h]$ at fixed $h(y)$ under the constraint (\ref{constraint}). By translational invariance, the minimum-energy configuration of the contact-line itself, i.e.\ the minimum of $E[h]$, is attained for a straight line $h(y)=h_0$ where $h_0$ denotes the equilibrium height.

In the next Section we compute the elastic energy of the contact line, i.e.\ $E[h]$ to second order in its deformations, i.e.\ $E[h] - E[h_0] = E_{\mathrm{el}}[h] + {\cal O}(h^3)$ with
\begin{equation}
E_{\mathrm{el}}[h] = \frac{1}{2} \int_q \epsilon_q h_{-q} h_q \ .
\end{equation}
Here $h(y) = \int_q h_q e^{i q y}$ and we denote $\int_q:= \int \frac{dq}{2 \pi}$. For a uniform deformation
$h(y)=h$ it takes the form:
\begin{equation}
E_{\mathrm{el}}[h] = \frac{1}{2} m^2 (h-h_0)^2 L_y \quad , \quad m^2 = \epsilon_{q=0}
\ ,
\end{equation}
which defines what we call the mass $m$, i.e.\ $m^2$ is the curvature of the parabolic well in which the
contact line sits because of gravity. We then calculate $\epsilon_q$, already announced in eq.\ (\ref{final1}).

\section{Contact-line elasticity}
\label{section3}
\subsection{Model in shifted coordinates}

We start by introducing a more convenient expression for the energy of the system. The constraint
$X> - h(y) \sin \varphi$ in the domain of integration is tedious to handle, so we introduce the function $z(x,y)$ as
\begin{eqnarray}
z(x,y) := Z(X = x - \tilde h(y) \sin \varphi,y)\ .
\end{eqnarray}
It satisfies  the same boundary condition $z(\infty,0)=0$. 
We have also defined
\begin{eqnarray}
\tilde h(y) := h(y) - h_0
\end{eqnarray}
i.e.\ $z$ is still the height along the vertical axis, but we have shifted the $X$ coordinate so
that the integration domain is
\begin{eqnarray}
x >  x_0 :=-h_0 \sin \varphi
\ .
\end{eqnarray}
Using that the derivatives, evaluated at $X=x- \tilde h(y) \sin \varphi$, satisfy
\begin{eqnarray}
 \partial_X Z(X,y)  &=&  \partial_x z(x,y) \\
 \partial_y Z(X,y)  &=& \partial_y z(x,y) + \tilde h'(y)  \sin \varphi ~ \partial_x z(x,y)\ , \qquad
\end{eqnarray}
one finds that the energy is now a functional noted $E[z,h] = \gamma {\cal E}[z,h]$ of $z(x,y)$ and $h(y)$ with
\begin{eqnarray} \label{genenergy}
\lefteqn{{\cal E}[z,h] = \int_{y} \rmd y \int_{x>x_0}\rmd x \bigg[  \frac{\kappa^2}{2} z(x,y)^2 -1 }\nonumber \\
&& +\sqrt{1 + [\partial_x z(x,y)]^2 + [\partial_y z(x,y) {+} \tilde h'(y) \sin \varphi \,\partial_x z(x,y)]^2 }  \bigg]\nonumber \\
&&\nonumber 
- \frac{\kappa^2}{6} \cos^2 \varphi\, \sin \varphi \int_y h(y)^3 \\
&& + (\sin \varphi  - \cos \theta) \int_y h(y) \ .
\end{eqnarray}
To derive this result, we have used the relation 
\begin{eqnarray}
\lefteqn{\int\limits_{X>-h(y) \sin \varphi} \!\!\!\!\!\!\rmd X  \sqrt{1+
[\partial_{X} Z]^{2}+[\partial_{y} Z]^{2} }-  \int\limits_{X>0} \rmd X \, \, 1 }\nonumber \\
&& = h(y) \sin \varphi +\int\limits_{X>-h(y) \sin \varphi} \!\!\!\!\!\!\rmd X  \sqrt{1+
[\partial_{X} Z]^{2}+[\partial_{y} Z]^{2} } -1\ ,\nonumber
\end{eqnarray}
which was then rewritten in terms of $x$. 

The new height function obeys the constraint
\begin{equation}  \label{boundary}
z(x_0,y) = h(y) \cos \varphi\ ,
\end{equation}
 i.e.\ it is specified on the edge $x=x_0$.

\subsection{Zero mode and calculation of the mass}

Let us first consider a uniform displacement of the line $h(y)=h$. We can then restrict to $y$-independent
height functions $z(x,y)=z(x)$, and the reduced energy takes the form \begin{eqnarray}
{ L_y^{-1} {\cal E}[z,h] }
& =& \int_{x>x_0} \rmd x\,  \bigg[ \sqrt{1 +  z'(x)^2 } -1  + \frac{\kappa^2}{2} z(x)^2 \bigg]\nonumber
\\
&& - \frac{\kappa^2}{6} h^3 \cos^2 \varphi \sin \varphi  + h (\sin \varphi - \cos \theta)\ . \qquad  
\end{eqnarray}
There are additional constraints at the boundary,
\begin{equation}\label{a7}
z(x=x_0)=h \cos \varphi\ ,
\end{equation}
and $z=0$ at infinity.
$h$ is arbitrary and not necessarily the preferred value at
equilibrium noted $h_0$. 

The linear variation of the reduced functional ${\cal E}[z,h]$ around an arbitrary configuration $(z(x),h)$, by
$(\delta z(x),\delta h)$ can be written upon integration by part as
\begin{eqnarray}
\lefteqn{\!\! \delta {\cal E}[z,h]/L_y} \nonumber \\
 & =& \int_{x>x_0} \rmd x\, \delta z(x) \bigg[\kappa^2 z(x) - \partial_x  \frac{z'(x)}{\sqrt{1+z'(x)^2}}  \bigg] \nonumber \\
&& +\, \delta z(x_0) \cos (\theta(x_0) + \varphi) - \frac{\kappa^2}{2}  \cos^2 \varphi \,\sin \varphi \, h^2 \delta h \nonumber \\
&& +\, \delta h (\sin \varphi - \cos \theta)\ . \label{var}
\end{eqnarray}
We have defined
\begin{eqnarray}\label{defcs}
 \cos (\theta(x) +\varphi)&=& - \frac{z'(x)}{\sqrt{1+z'(x)^2}} \\
  \sin (\theta(x) + \varphi) &=& \frac{1}{\sqrt{1+z'(x)^2}}\ ;
\end{eqnarray}
hence $\theta(x_0)$ is the contact angle at the wall and $\theta(x)+\varphi$ is the local angle
with respect to the vertical. Note that, because of $\kappa^2>0$, the profile decays and the
boundary term at infinity does not contribute. The constraint on the boundary
implies the additional relation
\begin{equation}\label{a77}
\delta z(x_0) = \delta h \cos \varphi \ .
\end{equation}
Consider now the function $z(x)=z_h(x)$ which minimizes the energy at fixed
$h$, i.e.\ $\delta h=0$ and $\delta z(x_0)=0$. This leaves the  variations 
in the bulk, given in the first line of (\ref{var}), leading to the stationarity condition
\begin{equation}\label{a6}
\kappa^2 z (x) = \frac{z'' (x)}{(1+z' (x)^{2})^{3/2}}\ .
\end{equation}
It can be  integrated once, 
\begin{equation}\label{a8}
\frac{\kappa^{2}}{2} z(x)^{2} =1  -\frac{1}{\sqrt{1+ z'(x)^{2}}}
\ ,
\end{equation}
where the integration constant was fixed by considering the limit of
$x\to \infty$. This yields at $x=x_0$, 
\begin{eqnarray}\label{a12a}\nonumber
 \frac{\kappa^{2}}2 {h^{2}} \cos^2 \varphi&=& 1 - \sin( \theta(x_0) + \varphi) \\
& =& 2 \sin^2\left(\frac{\theta(x_0)}{2} + \frac{\varphi}{2} - \pi/4\right)\ ,
\end{eqnarray}
using that $1-\sin x=2 \sin^2(x/2-\pi/4)$. 
The sign of the root $h \cos \varphi$ must be opposite to the sign of $\theta(x_0)+\varphi-\pi/2$,
and similarly when solving for $z$ in eq.~(\ref{a8}). This yields
\begin{eqnarray}\label{a12}
 h \cos \varphi &=& - 2 L_c \sin\left(\frac{1}{2} (\theta(x_0)+\varphi - \frac{\pi}{2})\right)
\ .
\end{eqnarray}
Integrating once more, we obtain the height profile $z=z_h(x)$ in the inverse form,
\begin{eqnarray}\label{a9}
 x (z) &=& x_ 0 + L_c \left[ \,\mbox{arcosh} \left(\frac{2 L_c}{z}\right) - \,\mbox{arcosh} \left(\frac{2 L_c}{h \cos \varphi}\right)\right] \nonumber  \\
&& - L_{c} \left( \sqrt{4-\frac{z^2}{L_{c}^2}} - \sqrt{4-\frac{h^2 \cos^2 \varphi}{L_c^2}} \right)\ .
\end{eqnarray}
The integration constant was chosen to satisfy the constraint (\ref{a7}).

The reduced energy of the uniform deformation $h(y)=h$ 
is thus a simple function ${\cal E}(h)={\cal E}[z_h,h]$, whose derivative, ${\cal E}'(h)$ is easy to
obtain. Indeed we can use the general variational formula (\ref{var})
around $(z_h(x),h)$ setting $\delta z(x)=\delta h \partial_h z_h(x)$ with
$\partial_h z_h(x_0)=1$ from the constraint (\ref{a7}). The bulk contribution
is zero, due to the ``equation of motion'' (\ref{a6}) satisfied by $z_h(x)$.  The boundary
contributions, i.e.\ the second and third lines in (\ref{var}),  give
\begin{eqnarray}\label{a12f}
 {\cal E}'(h)/L_y &=& \cos \varphi \cos(\theta(x_0)+\varphi) \nonumber \\
&& - \frac{\kappa^2}{2} h^2 \cos^2 \varphi \sin \varphi  + \sin \varphi - \cos \theta
\ . \qquad 
\end{eqnarray}
Using eq.~(\ref{a12a}) this simplifies to 
\begin{eqnarray}  \label{deriv} 
 {\cal E}'(h)/L_y = \cos \theta(x_0) - \cos \theta \ ,
\end{eqnarray}
where $\theta(x_0)$ is an implicit function of $h$, using (\ref{a12}). We first note that this allows to recover the usual condition for equilibrium, ${\cal E}'(h_0)= 0$ namely that the
local contact angle $\theta(x_0) = \theta$, the equilibrium contact angle. The equation determining
$h_0$ is thus (\ref{a12}), with $\theta(x_0)= \theta$, 
\begin{equation}\label{a12b}
 h_0 \cos \varphi = - 2 L_c \sin\left(\frac{1}{2} (\theta +\varphi - \frac{\pi}{2})\right)\ .
\end{equation}
Inverting eq.~(\ref{a12b}) one finds from (\ref{deriv}) away from equilibrium
\begin{eqnarray}\label{a12c}
 \frac{{\cal E}'(h)}{L_y} &=& h \,\kappa \cos^2 \varphi \,\sqrt{1 - \frac{1}{4} \kappa^2 h^2 \cos^2 \varphi }\nonumber \\
&& +  \left( 1 - 
\frac{1}{2} h^2 \kappa^2 \cos^2 \varphi \right) \sin \varphi - \cos \theta
\ . \qquad
\end{eqnarray}
The mass for the zero mode is thus, after various simplifications
\begin{eqnarray}\label{a12d}
 m^2 = \frac{\gamma {\cal E}''(h_0)}{L_y} = \frac{\gamma \kappa \sqrt{2} \cos \varphi \sin \theta}{\sqrt{1+\sin(\theta + \varphi)}} 
\ .
\end{eqnarray}
An interesting special case is that of a flat interface, $\theta +\varphi =\frac{\pi}{2}$.  Then 
\begin{equation}\label{k2}
 m^2 =  \gamma \kappa \sin^2(\theta)\ .
\end{equation}

\begin{figure}
\Fig{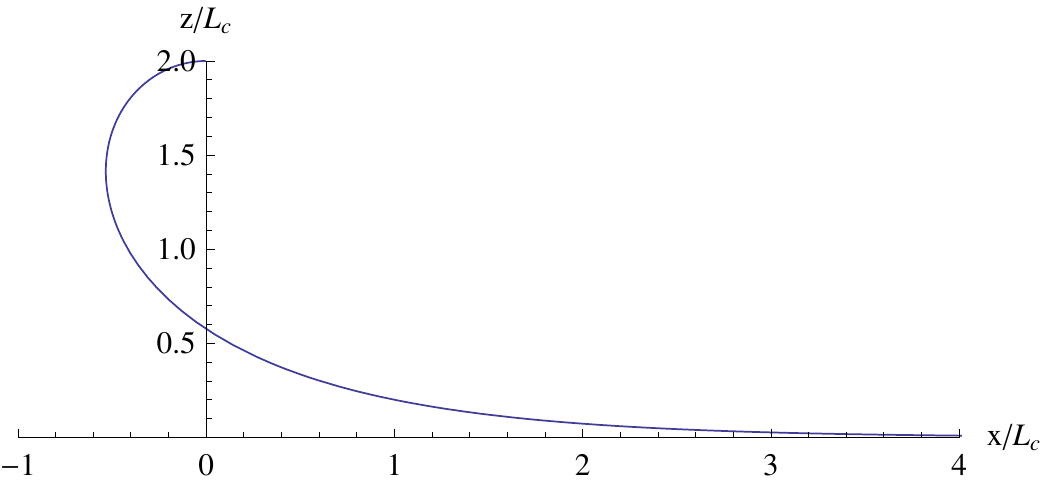}
\caption{The height profile $z (x)$, in units of the capillary length
$L_{c}$, for $\theta =30^{\circ}$ and $\varphi=0$. Different 
contact-angles $0<\theta<\pi/2$ are obtained by moving the graph
left/right. The case $\pi/2<\theta<\pi$ is obtained by the reflection $z \to -z$.}
\label{a4}       
\end{figure}

\subsection{Elastic energy for arbitrary deformations}

We now go back to the general case for $h(y)$. The reduced energy functional ${\cal E}[z,h]$ given in (\ref{genenergy})
explicitly depends on $h(y)$. In addition $h(y)$ also enters in the boundary condition (\ref{boundary}). 
It is important to distinguish these two dependences.

We call $z_h(x,y)$ the profile which minimizes the reduced energy functional ${\cal E}[z,h]$ with fixed $h(y)$,
hence with respect to bulk variations $\delta z(x,y)$ only, 
\begin{eqnarray}\label{147}
 \frac{\delta {\cal E}}{\delta z(x,y)|_{\mathrm{bulk}}}[z_h,h] = 0 \ ,
\end{eqnarray}
i.e.\ a partial (functional) derivative w.r.t. $z$ only. 

We now expand around the equilibrium solution as follows:
\begin{eqnarray} \label{defzt}
 h(y) &=& h_0 + \tilde h(y) \\
 z_h(x,y) &=& z_0(x) + \tilde z(x,y) \nonumber 
\ ,
\end{eqnarray}
where $z_0(x):=z_{h_0}(x)$ is the equilibrium profile determined in the previous section. Since $\tilde z$ is 
of order ${\cal O}(\tilde h)$, to compute the elastic energy we need to expand the minimum energy to second order in $\tilde h$ and $\tilde z$.
Expanding (\ref{genenergy}) to second order in the {\it explicit} dependence on $\tilde z$ and $\tilde h$ we have
\begin{eqnarray}
&& {\cal E}[z_{h},h] - {\cal E}[z_0,h_0] = \delta^{(1)} {\cal E} + \delta^{(2)} {\cal E} + {\cal O}(\tilde h^3) \\
&& \delta^{(2)} {\cal E} = \delta^{(2,1)} {\cal E} + \delta^{(2,2)} {\cal E} + \delta^{(2,3)} {\cal E} \nonumber \ .
\end{eqnarray}
The first term is the linear variation, 
\begin{eqnarray}
 \delta^{(1)} {\cal E} = \frac{\delta E}{\delta z}[z_0,h_0] \cdot \tilde z +  \frac{\delta E}{\delta h}[z_0,h_0] \cdot \tilde h\ ,
\end{eqnarray}
whereas $\delta^{(2,p)} {\cal E}$ are second order variations computed below. Let us start with $\delta^{(1)} {\cal E}$:
\begin{eqnarray}
\delta^{(1)} {\cal E}  &=& \int_y \int_{x>x_0}  \tilde z(x,y) \left[\kappa^2 z_0(x) - \partial_x  \frac{z_0'(x)}{\sqrt{1+z_0'(x)^2}}  \right] \nonumber \\
&& + \int_y \delta z(x_0,y) \cos (\theta(x_0) + \varphi) \nonumber \\&& - \frac{\kappa^2}{2}  \cos^2 \varphi \,\sin  \varphi\, h_0^2 \int_y \tilde h(y) \nonumber \\
&& + \int_y \tilde h(y) (\sin \varphi - \cos \theta) \label{var2}
\ .
\end{eqnarray}
The first line is the bulk variation which vanishes since $z_0(x)$ satisfies the stationarity condition (\ref{a6}). The rest
is the sum of the {\it boundary} variation in $z$ and the variation in $h$. Thanks to the exact constraint
\begin{equation} \label{bc}
\tilde z(x_0,y) = \tilde h(y) \cos \varphi\ , 
\end{equation}
one can check, using the results of the previous section, that this sum vanishes identically if $h_0$ is the equilibrium value. Hence we 
have $\delta^{(1)} {\cal E}=0$. Since $\tilde z$ also contains subdominant  $\tilde h^2$ contributions, this is a
quite useful observation, as we do not need to worry about them in computing the elastic energy. They drop out by virtue of the equilibrium condition. 

Let us now study the three second variations. The first one is
\begin{eqnarray} \label{e21}
&&\!\!\! \delta^{(2,1)} {\cal E} =  \frac{1}{2} \tilde z \cdot \frac{\delta^2 E}{\delta z \delta z}[z_0,0] \cdot \tilde z \\
&&\!\!\! =  \frac{1}{2} \int\limits_{y,x>x_0}  \kappa^2 \tilde z(x,y)^2 + \frac{[\partial_x \tilde z(x,y)]^2}{[1+z_0'(x)^2]^{3/2}}  + \frac{[\partial_y \tilde z(x,y)]^2}{[1+z_0'(x)^2]^{1/2}} \nonumber\ .
\end{eqnarray} 
The second one is
\begin{eqnarray}  \label{e22}
 \delta^{(2,2)} {\cal E} &=& \tilde h \cdot \frac{\delta^2 E}{\delta h \delta z}[z_0,h_0] \cdot \tilde z \\
& =& \sin \varphi \int_{y, x> x_0} h'(y) \frac{z'_0(x)}{\sqrt{1+ z'_0(x)^2}} \partial_y \tilde z(x,y) \nonumber\ ,
\end{eqnarray}
where we have used $\tilde h'(y)=h'(y)$ to alleviate the notation. 
The third contribution is
\begin{eqnarray}
\delta^{(2,3)} {\cal E}   &=& \frac{1}{2} \tilde h \cdot \frac{\delta^2 E}{\delta h \delta h}[z_0,h_0] \cdot \tilde h \nonumber \\
& =& \delta^{(2,3a)} {\cal E} + \delta^{(2,3b)} {\cal E}\ ,
\end{eqnarray}
where we write separately the second derivative with respect to the explicit 
dependence on $\tilde h(y)$ of the surface energy, namely
\begin{eqnarray} \label{3a}
 \delta^{(2,3a)} {\cal E} = \frac{1}{2} \sin^2 \varphi \int_{y}\int_{x>x_{0}} h' (y)^{2}\frac{z_0'
(x)^{2}}{\sqrt{1+z_0' (x)^{2}}}\ , ~~~
\end{eqnarray}
and of the gravitational one, 
\begin{eqnarray} \label{3b}
 \delta^{(2,3b)} {\cal E} = - \frac{1}{2} \kappa^2 \cos^2 \varphi\, \sin \varphi \,h_0 \int_{y} \tilde h(y)^2\ .
\end{eqnarray}
To compute the first two contributions one needs to specify the properties of $\tilde z(x,y)$
which follow from its definition (\ref{defzt}). The function $z_h(x,y)$ must obey equation (\ref{147}), for any $h$. 
Hence we can expand this equation order by order in $h$. For $h=0$ it yields again the
stationarity condition ({\ref{a6}) for $z_0(x)$; to first order it yields
\begin{equation}\label{147b}
 \tilde z \cdot \frac{\delta {\cal E}}{\delta z \delta z(x,y)|_{\mathrm{bulk}}}[z_0,h_0] + \tilde h \cdot \frac{\delta {\cal E}}{\delta h \delta z(x,y)|_{\mathrm{bulk}}}[z_0,h_0]  = 0 \nonumber 
\ ,
\end{equation}
which, in explicit form, yields the equation obeyed by $\tilde z(x,y)$ (to linear order in $h$ which is all we need here):
\begin{eqnarray}\label{172}
&&\!\!\!\!\!\!\!{\left[ \kappa^2 -  \partial_x \frac{1}{[1+ z_0'(x)^2 ]^{3/2}}  \partial_x -  \frac{1}{[1+ z_0'(x)^2 ]^{1/2}}  \partial_y^2 \right]  \tilde z(x,y)} \nonumber \\
&&\ = \sin \varphi\, h''(y) \frac{z'_0(x)}{[1+ z_0'(x)^2 ]^{1/2}}  \ .
\end{eqnarray} 
It must be solved with the boundary condition (\ref{bc}) and $\tilde z(x=\infty,y)=0$. 

Remarkably, this complicated looking equation, which depends on the profile $z_0(x)$ known
only in the implicit form (\ref{a9}), can be solved analytically. First of all, using the stationarity equation
(\ref{a6}) satisfied by $z_0(x)$, one notes that
\begin{eqnarray} \label{46}
 \tilde z(x,y) = \tilde z_1(x,y) := - \sin \varphi ~ h(y) z'_0(x)
\end{eqnarray} 
is a particular solution of (\ref{172}) which vanishes at $x=\infty$ and
takes the value 
\begin{equation} \label{z1}
\tilde z_1(x_0,y)=  \sin \varphi \cot(\theta+\varphi) h(y)
\end{equation}
at the boundary. The full solution can thus be
written as
\begin{eqnarray}
 \tilde z(x,y) = \tilde z_1(x,y) + \tilde z_2(x,y) \ ,
\end{eqnarray} 
where $\tilde z_2(x,y)$ satisfies the homogeneous equation, i.e.
(\ref{172}) setting the r.h.s. to zero. The  boundary condition (\ref{boundary}) implies
\begin{eqnarray} \label{bc2}
 \tilde z_2(x_0,y) = \frac{\sin \theta}{\sin(\theta+\varphi)} h(y) \ . 
\end{eqnarray} 
Both $\tilde z_1$ and $\tilde z_2$   vanish at $x=\infty$. 

To solve the homogeneous equation, 
we go to Fourier space in $y$-direction,  and introduce a new variable in $x$-direction.  We thus look for the solution in the
form
\begin{eqnarray} \label{z2}
 \tilde z_2(x,y) &=& \frac{\sin \theta}{\sin(\theta+\varphi)} \int_q e^{i q y} \tilde h(q) F_{\tilde q}(S(x))\qquad  \\
 S(x) &=& \sin(\theta(x) + \varphi) = \frac{1}{\sqrt{1+ z'_0(x)^2} }\label{51} \\
 \tilde q &=& q/\kappa
\end{eqnarray} 
with $S(x_0)=\sin(\theta+\varphi)$. 
Deriving eq.~(\ref{51}) w.r.t.\ $x$, and using (\ref{a6}) and (\ref{a8}) to express the result in terms of $S(x)$, we obtain the rule for changing the derivatives, 
\begin{equation}\label{53}
\partial_{x} = \kappa \frac{\sqrt{2} (1-S) \sqrt{S+1}}{S}
\partial_{S}\ .
\end{equation}
The resulting equation for the function $F_{\tilde q}(S)$ reads
\begin{eqnarray}\label{184}
  &&(\tilde q^2 S+1)  F_{\tilde q}(S)+(1-S) (7 S^2+S-4)
   F_{\tilde q}'(S) \nonumber \\
   &&\qquad  -2 (S-1)^2 S (S+1) F_{\tilde q}''(S) = 0\ ,
\end{eqnarray}
where $\sin(\theta+\varphi) < S < 1$. The constraint (\ref{bc2}) implies the boundary conditions 
$F_{\tilde q}(\sin(\theta+\varphi))=1$ and $F_{\tilde q}(1)=0$. After some search, the general solution of
eq.~(\ref{184}) which satisfies $F_{\tilde q}(1)=0$ is found to be 
\begin{eqnarray}
 F_{\tilde q} (S) &=& \frac{g_r(S)}{g_r(\sin(\theta+\varphi))} \\
 g_r(S) &=& \frac{(1-S)^{r/2}}S  \left(1+\frac{\sqrt{S+1}}{\sqrt{2}}\right)^{\!\!-r} \nonumber \\
&& \times \left(2 r^2+3 \sqrt{2} r \sqrt{S+1}+3 S+1\right)\ ,
\end{eqnarray}
where we denoted
\begin{eqnarray}
 r := \sqrt{1 + \tilde q^2} \ .
\end{eqnarray}

We now evaluate the second variation. Consider first the sum of (\ref{e21}) and (\ref{e22}).
Using the equation of motion (\ref{172}) for $\tilde z$, the combination $\delta^{(2,1)} {\cal E} +  \frac12 \delta^{(2,2)} {\cal E}$
can be integrated by part.  Therefore, we obtain for the combination $[\delta^{(2,1)} {\cal E} +  \frac12 \delta^{(2,2)} {\cal E}]+\frac12 \delta^{(2,2)} {\cal E}$
\begin{eqnarray} \label{new}
&& \delta^{(2,1)} {\cal E} +  \delta^{(2,2)} {\cal E} =  - \frac{1}{2}\int_{y}\frac{ \tilde z(x_0,y)}{[1+z_{0}'
(x)^{2}]^{3/2}} \partial_{x}  \tilde z(x,y)\Big|_{x=x_0} \nonumber \\
&& ~- \frac{1}{2} 
\sin \varphi \int_{y, x> x_0} h''(y) \frac{z'_0(x)}{\sqrt{1+ z'_0(x)^2}} \tilde z(x,y)\ .
\end{eqnarray}
The last term has been integrated by part w.r.t.\ $y$ \footnote{Note that the integration by part of $\partial_y$ produces no surface term. This can be
made rigorous considering a periodic modulation $h(y)$. We thus restrict here to functions that can
be written as sum of periodic modulations, or have compact support.}. 
To continue, we  note the useful equality 
\begin{eqnarray}\label{equal}\nn
&& - \int_{y}\frac{ \tilde z_{1} (x_0,y)}{[1+z_{0}'
(x)^{2}]^{3/2}} \partial_{x}  \tilde z_{2} (x,y)\Big|_{x=x_0} 
\\ \nn
&& \qquad = - \int_{y}\frac{ \tilde z_{2} (x_0,y)}{[1+z_{0}'
(x)^{2}]^{3/2}} \partial_{x}  \tilde z_{1} (x,y)\Big|_{x=x_0}
\\
&& \qquad\hphantom{=}+ \sin \varphi \int_{y, x> x_0} h''(y) \frac{z'_0(x)}{\sqrt{1+ z'_0(x)^2}} \tilde z_{2}(x,y) \ ,\qquad ~~
\end{eqnarray}
which is a consequence of the two different ways to integrate by part
\begin{eqnarray}\label{yy1}
&&  \int_{y,x>x_0}  \bigg[\kappa^2 \tilde z_{1}(x,y)
\tilde z_{2}(x,y) + \frac{[\partial_x \tilde z_{2}(x,y)][\partial_x 
\tilde z_{1}(x,y)]}{[1+ z_0'(x)^2]^{3/2}}  \nonumber \\
&& \qquad \qquad + \frac{[\partial_y 
\tilde z_{1}(x,y)][\partial_y  \tilde z_{2}(x,y)]}{[1+z_0'(x)^2]^{1/2}} \bigg]\nonumber \ ,
\end{eqnarray}
and to use the equation of motion for $\tilde z_1$ (inhomogeneous) and $\tilde z_2$ 
(homogeneous). 

Inserting $\tilde z=\tilde z_1+\tilde z_2$ into (\ref{new}) and using the equality (\ref{equal}) we get
\begin{eqnarray} \label{tot}
&& \!\!\!\delta^{(2,1)} {\cal E} +  \delta^{(2,2)} {\cal E} = - \frac{1}{2}\int_{y}\frac{ \tilde z_2(x_0,y)}{[1+z_{0}'
(x)^{2}]^{3/2}} \partial_{x}  \tilde z_2(x,y)\Big|_{x=x_0} \nonumber \\
&& - \frac{1}{2}\int_{y}\frac{ \tilde z_1(x_0,y)}{[1+z_{0}'
(x)^{2}]^{3/2}} \partial_{x}  \tilde z_1(x,y)\Big|_{x=x_0} \nonumber \\
&&  - \int_{y}\frac{ \tilde z_2(x_0,y)}{[1+z_{0}'
(x)^{2}]^{3/2}} \partial_{x}  \tilde z_1(x,y)\Big|_{x=x_0} \nonumber \\
&& - \frac{1}{2} 
\sin \varphi \int_{y, x> x_0} h''(y) \frac{z'_0(x)}{\sqrt{1+ z'_0(x)^2}} \tilde z_1(x,y)\ .
\end{eqnarray}
We now discuss simplifications. Firstly, the last term in eq.\ (\ref{tot}) exactly  cancels $\delta {\cal E}^{(2,3b)}$; this is shown using (\ref{46}).  

Secondly, from (\ref{46}), (\ref{a6}) and (\ref{boundary}), we obtain   
\begin{eqnarray}
 \frac{\partial_{x}  \tilde z_1(x,y)}{[1+z_{0}'(x)^{2}]^{3/2}}\Big|_{x=x_0} 
 = - \sin \varphi \cos \varphi \kappa^2 h(y) h_0 \ .
\end{eqnarray}
This shows that the second line, half the third line and $\delta^{(2,3b)} {\cal E}$ cancel; the remaining half of the third line gives the first term reported in eq.~(\ref{63}) below. The second term comes from the first line of (\ref{tot}), using (\ref{z2}), so we get finally:
\begin{eqnarray}\label{63}
\lefteqn{ \!\!\!\delta^{(2)} {\cal E} = \frac{\sin \varphi \cos \varphi \sin \theta}{2 \sin(\theta+\varphi)} \kappa^2 \,h_0 
\int_y h(y)^2}  \\
&&- \frac{1}{2} \sin^2 \theta \sin(\theta+\varphi) 
\int_q h_{q}h_{-q}  F_{\tilde q}(S(x)) \partial_x F_{\tilde q}(S(x))\Big|_{x=x_0}\nonumber
\ .
\end{eqnarray}
To compute the second term we use rule (\ref{53}), where at the end $S$ must be evaluated on the boundary $S=\sin(\theta+\varphi)$. 
To compute the first term we use the value (\ref{a12a}) for $h_0$.

This yields our final result for the elastic energy,
\begin{equation}\label{yy2}
 E_{\mathrm{el}}[h] = \frac{1}{2}\int_{q} \epsilon _{q}  h_{q} h_{-q}\ ,  
\end{equation}
with 
\begin{eqnarray}\label{final}
  \frac{\epsilon_{q}}{\kappa \gamma} &=& \frac{ \sin (\theta ) \cos (\varphi )}{
 t} + \frac{\left(r^2-1\right)\left [t
 (r+t)+1\right]\sin ^2(\theta )}{ t \left(r^2+3 r t+3 t^2-1\right)}  \nonumber \\
   t&=& \sqrt{\frac{\sin(\theta +\varphi )+1}{2}} \quad , \quad  r= \sqrt{1+\frac{q^{2}}{\kappa^{2}}} \ .
\end{eqnarray}
One finds that $\epsilon_q$ is a scaling function of $q/\kappa$ which reproduces formula (\ref{a12d}) for 
the energy of a uniform mode $\epsilon_{q=0}= m^2$ as computed in the previous section,
and which behaves as $\epsilon_{q} \approx \kappa \gamma \sin^2 \theta |q|$ for large $|q|$.


When 
$\varphi+\theta=\pi/2$, the equilibrium shape of the interface is
flat. Thus the elastic energy is expected to simplify. Indeed, it
becomes 
\begin{equation}\label{f1}
 \frac{\epsilon_{q}}{\kappa \gamma} = \sin ^2(\theta )
 \sqrt{1+\frac{q^{2}}{\kappa^{2}}}\ .
\end{equation}

\begin{figure}
\Fig{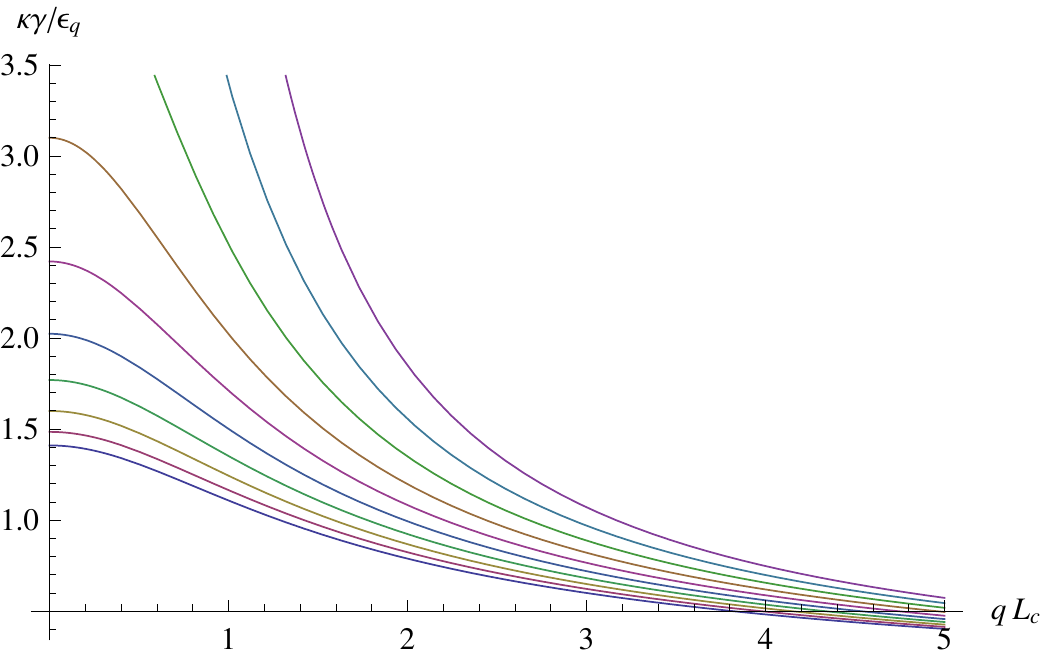}
\caption{$\kappa\gamma/\epsilon_q$ as a function of $q L_c$, for $\theta=40^\circ$, $\varphi=0^\circ$, $10^\circ, \ldots,90^\circ$ (from bottom to top).}
\end{figure}

\subsection{The contact-line profile as a means of measuring
$\epsilon_{q}$}\label{s:measure_epsilon}
Suppose the contact line is in force-free equilibrium. Then pull on it
with a force per unit length:
\begin{equation}\label{f2}
 F(y) = \frac{f}{2\delta} \theta (|y|<\delta)
\ .
\end{equation}
This leads in linear response (i.e.\ for the quadratic in $h$ elastic
energy we are using) to the following profile 
\begin{equation}\label{f3}
h (y) =f \int_{0 }^{\infty }\frac{dq}{\pi }\frac{\cos
( qy)}{\epsilon_{q}} \frac{\sin ( q\delta)}{q\delta} \ .
\end{equation}
The limit of a $\delta$-like force is recovered in the limit of
$\delta \to 0$, which eliminates the last factor of $\sin(q\delta)/
(q\delta)$. However the latter makes the integral convergent at large $q$. We have plotted for $\theta =40^{\circ
}$ and $\delta =0.3$ two solutions on figure \ref{f:profile}, one for  $\varphi =0^{\circ}$
(bottom) and the other for $\varphi =45^{\circ}$ (top). 
\begin{figure}
\fig{1}{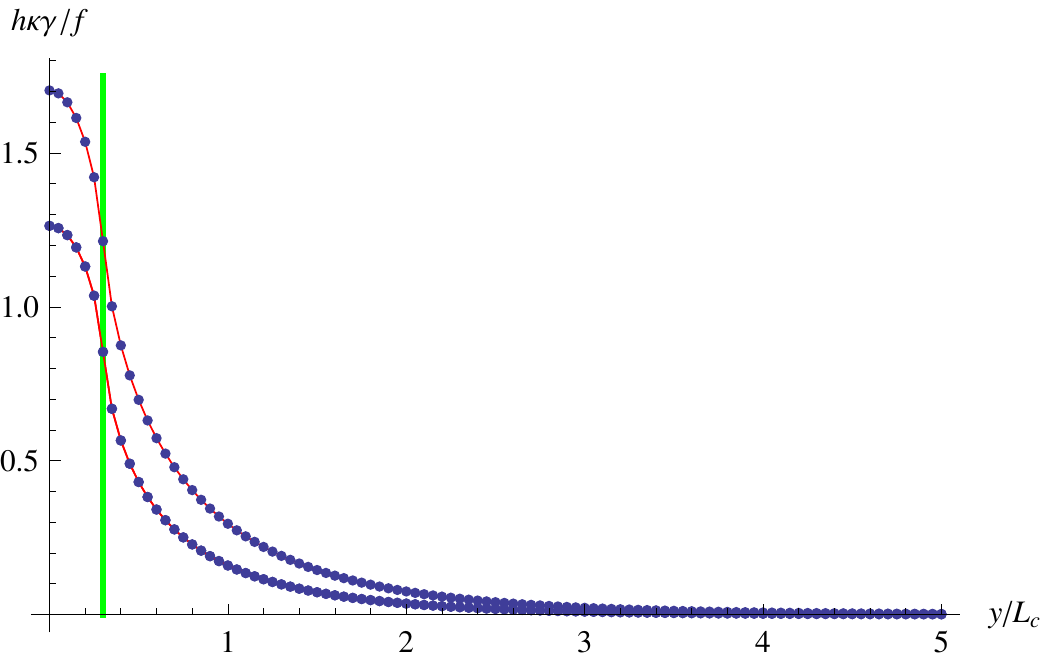}
\caption{Sample profiles $h(y)$, in units of $f/(\kappa \gamma)$, 
plotted as a function of $y/L_c$, 
for $\theta =40^{\circ }$, $\varphi =0^{\circ}$
(bottom) and $\varphi =45^{\circ}$ (top). The profiles are the response
to a force inside the green box $|y|<\delta=0.3 L_c$. The profile has to
be continued symmetrically to the left.}
\label{f:profile}
\end{figure}
\begin{figure}
\fig{1}{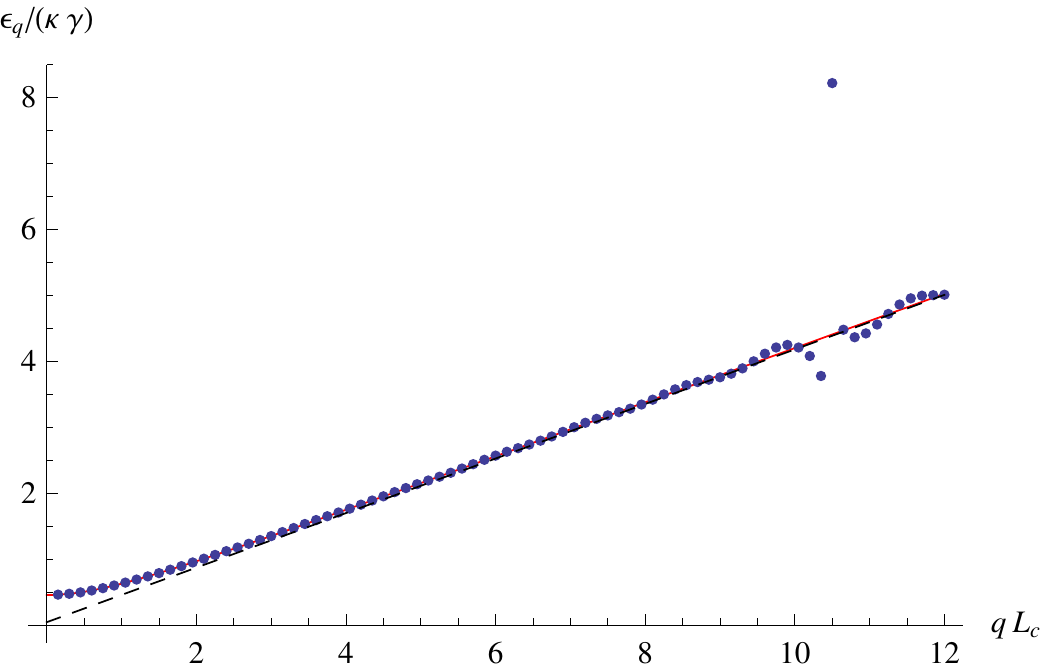}
\fig{1}{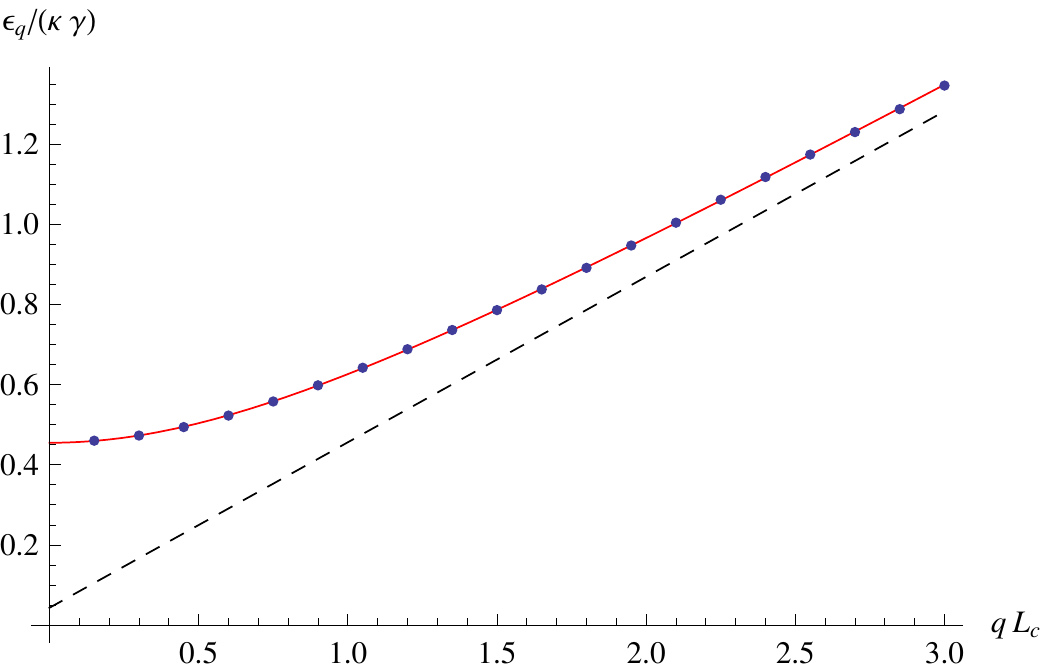}
\caption{Top: Blue points: Reconstruction of $\epsilon_{q}$, in units of 
$\kappa \gamma$, as a function of $q/\kappa$, using (\ref{reconstruct})
from the points of the bottom curve of figure \ref{f:profile}, i.e.\ a
vertical plate geometry $\varphi=0$. Solid red
line: the analytical result. One sees a  numerical problem
appearing at large $q$, at $q  = \pi/\delta $, where $\delta$ is the
box-size in figure \ref{f:profile}. However this is already far in the
linear asymptotic regime (dashed line). Bottom: The same plot for
smaller $q$. One sees that $\epsilon_{q}$ is well reconstructed.}
\label{f:reconstruct}
\end{figure}
We  note that  $\epsilon_{q}$ can be reconstructed from the profile as
follows 
\begin{equation}\label{reconstruct}
\epsilon_{q} = f \frac{\sin(q\delta)}{q \delta} \left[2{
\int_{0}^{\infty} \rmd y \cos(qy)h(y)} \right]^{-1}
\ .
\end{equation}
A reconstruction of $\epsilon_{q}$ starting from the blue points on
the top of figure \ref{f:profile} is given in figure \ref{f:reconstruct}.

\section{Avalanche-size distributions}
\label{section4}
\subsection{The model}\label{k1}
\begin{figure}[b]
\centerline{\Fig{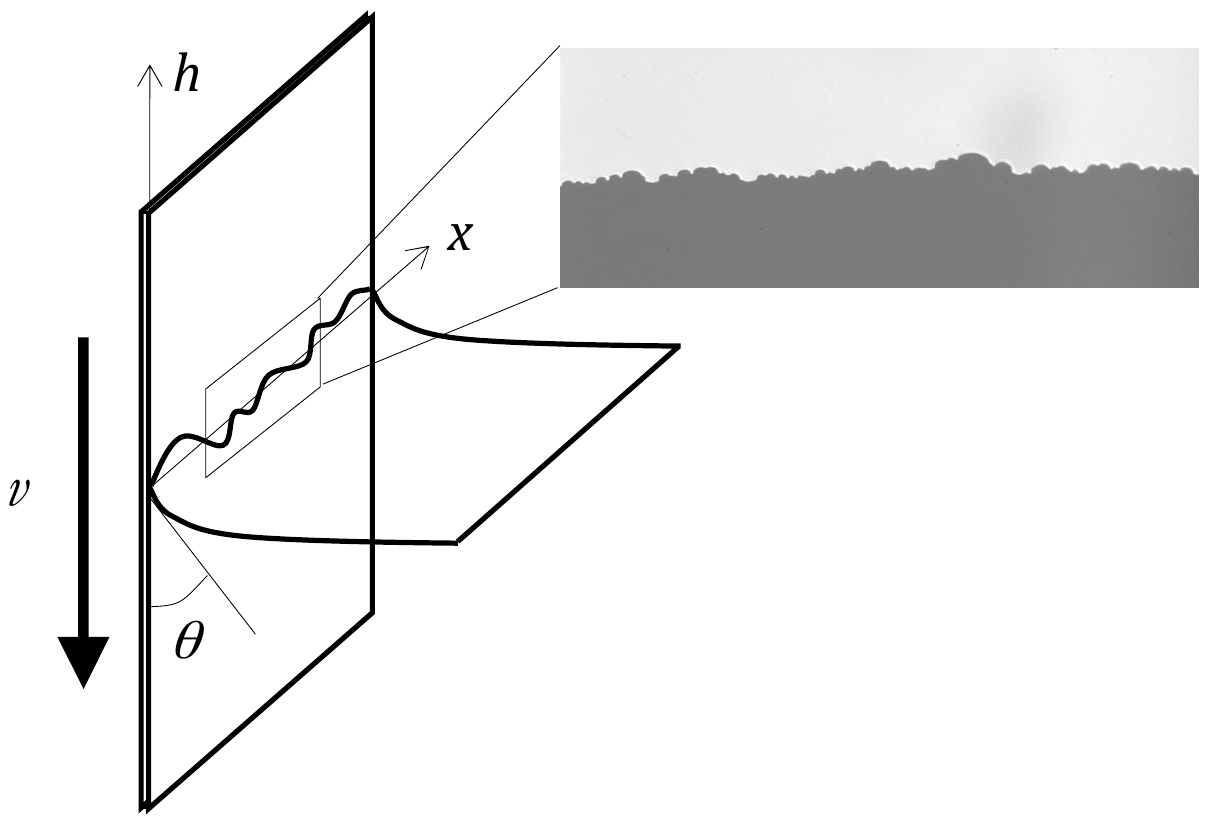}}
  \caption{Sketch of the experimental setup used in \cite{LeDoussalWieseMoulinetRolley2009}. The size of the image in the inset is 1.5 mm.
  }\label{setup}
\end{figure}

We now study the case of a disordered plate, which is immersed with
velocity $v$ into the liquid reservoir. This is the geometry of the experiment described in Ref. \cite{LeDoussalWieseMoulinetRolley2009}. We use some of the notations defined there:
$x$ denotes the coordinate along the contact line (denoted $y$ in the previous section) and
$u(x)$ the height of the contact line along the plate, in the frame of the plate. We reserve the notation $h(x)$ to the contact-line height measured in the laboratory frame, with the relation\footnote{In the experiment of Ref.~\cite{LeDoussalWieseMoulinetRolley2009}, there is an additional slowly varying offset.}:
\begin{equation}\label{k2}
 h(x,t) = u(x,t) - w \  , \qquad  w = vt \ .
\end{equation}
Even though it is a dynamical problem, it is useful to introduce the energy
\begin{equation}\label{a1}
{\cal H}[u] = \int_{0}^{L} \rmd x\, \frac{m^2}{2} \left[u(x)-w\right]^2  + V (x,u(x)) + \delta{E}[u]\ .
\end{equation}
Here $m^{2}$  is the mass of the zero mode
$q=0$, and $\delta E [u]:= \frac{1}{2}\int_{q}
[\epsilon _{q}-\epsilon _{0}]  u_{q} u_{-q}  $ the remaining part of the elastic energy at
non-zero wave vector $q\neq 0$. For the contact-line, 
$m^{2}=  \epsilon_{0}$ and $\delta E[u]$ are both  given in (\ref{final}). We neglect possible non-linear elastic terms \cite{LeDoussalWieseRaphaelGolestanian2004,BachasLeDoussalWiese2006}. The function $V (x,u)$ is a random potential, whose derivative
$\partial_{u}V (x,u)$ is short-ranged correlated, modeling the disordered substrate. 

The contact line is pinned by the disorder, but also trapped in the quadratic well with
curvature $m^{2}$. Advancing the well-position $w$ by immersing the plate  with
velocity $v$ leads to a motion of the contact line, which we now study in the quasi-static limit, i.e.\ the limit of small $v$. From the energy form (\ref{a1}) one can derive, upon various assumptions about the fluid,  equations of motion (see e.g.\ \cite{NikolayevBeysens2003,GolestanianRaphael2001} and references therein). Our assumption here is that in the quasi-static limit, they lead to the same statistics as the simplest over-damped model studied in Ref. \cite{LeDoussalWiese2008c}, at least to the level of approximation
that we use here (i.e.\ lowest order, i.e.\ one-loop FRG). Furthermore,
although quasi-static dynamics and pure statics are different, they lead, in the same order of approximation, to identical rescaled avalanche-size distributions. Deviations are expected only at the next, i.e.\ two-loop order. This justifies our studying of the energy form (\ref{a1}), and, as we will see, our point is that noticeable effects already arise from the precise form of the elastic kernel. 

\subsection{Global statistics of avalanches}

\subsubsection{Definitions}

\begin{figure}[t]
\centerline{\fig{0.75}{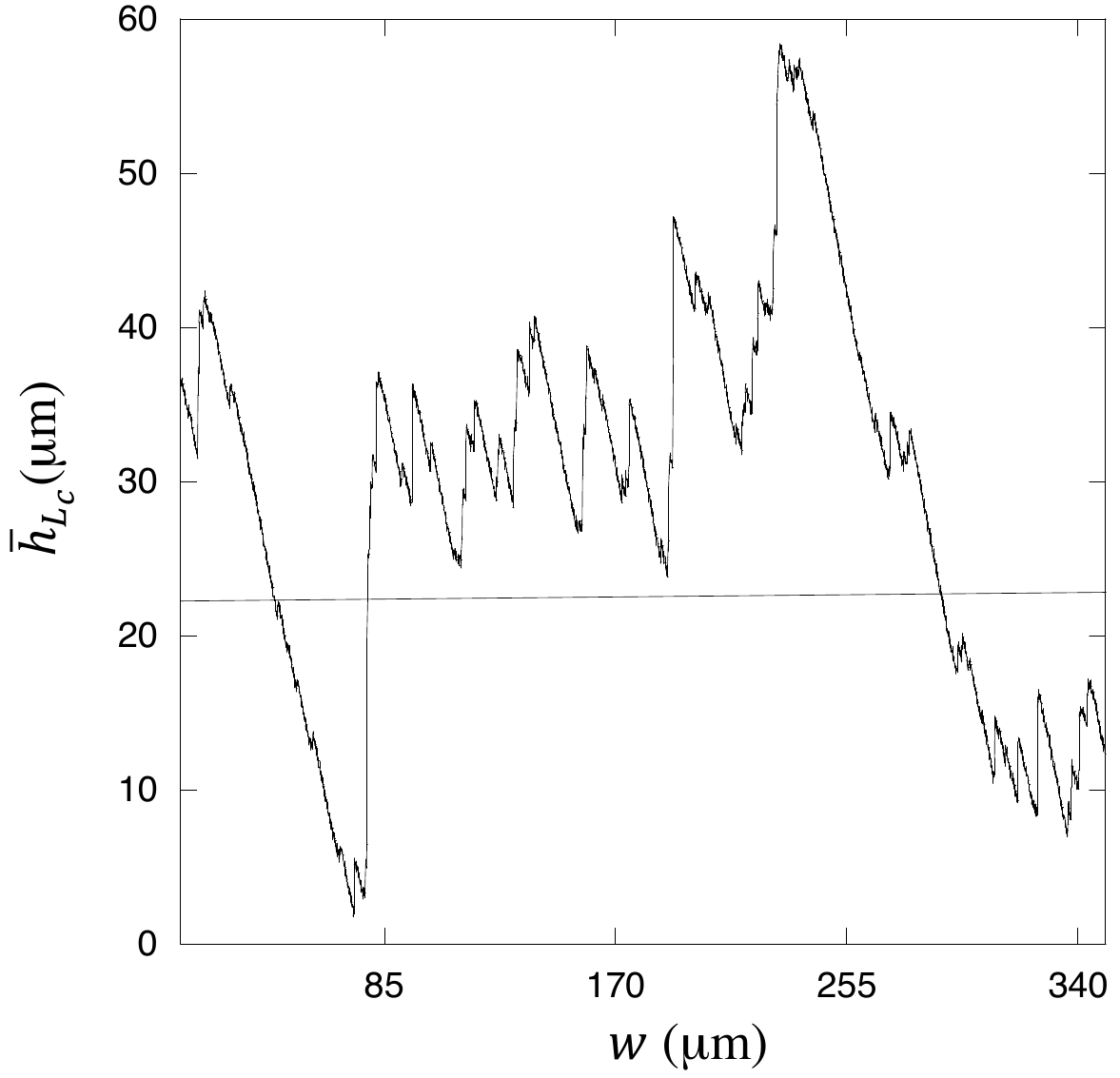}}
  \caption{ Data from \cite{LeDoussalWieseMoulinetRolley2009}: Height of the contact line  $\bar{h}(w)$ averaged
  over $2L_c$, as a function of the position $w$ of the plate (system:
  iso/Si). The fast depinning events (upwards) are clearly
  visible. Between them, the contact line moves downwards at the plate
  velocity $v$ (here 1 $\mathrm{ \mu m/s}$). The straight line is the
  reference level $h_0$.
  }\label{h(t)}
\end{figure}
As can be seen on figure \ref{h(t)},
the experiment \cite{LeDoussalWieseMoulinetRolley2009}  shows, as
predicted by the theory \cite{LeDoussalMiddletonWiese2008,LeDoussalWiese2008c}, that the motion of the contact-line proceeds by 
sudden jumps, i.e.\ avalanche motion.
For an avalanche occurring at a given position $w$ of the center of the quadratic well,
one defines $u_{w}^{-} (x)$ the contact-line position just before the avalanche
and $u_{w}^{+} (x)$ its position just after. The size of the avalanche is
defined as
\begin{equation}\label{k3}
S:= \int_{0}^{L}\rmd x\,  \left[u_{w}^{+} (x)-u_{w}^{-} (x)\right] = \int_{0}^{L}\rmd x\,  \left[h_{w}^{+} (x)-h_{w}^{-} (x)\right]\ .
\end{equation}
Hence one can measure the avalanche size $S$ as the area swiped
both in the $u$ or in the $h$ coordinate system.

We now recall the main results of
Refs.~\cite{LeDoussalMiddletonWiese2008,LeDoussalWiese2008c}  and
apply the general formulation to various cases, including the
 contact line for which we have computed the elasticity in section 3.
The characteristic rescaled function $\tilde Z(\lambda)$ is defined as:
\begin{eqnarray}\label{69}
 \tilde Z(\lambda)&:=& \frac{S_m}{ \left< S \right>} \left< e^{\lambda S/S_m} - 1 \right>  \\
 S_m &:=& \frac{ \left< S^2 \right>}{2  \left< S \right> } = \frac{- \Delta'(0^+)}{m^4} \label{70}
\ .
\end{eqnarray}
All averages $\langle \ldots \rangle$ are over the normalized probability density $P(S)$ of avalanches. By definition
$\tilde Z (\lambda) = \lambda +\lambda^{2}+ {\cal O}(\lambda^{3})$. The function $\Delta(w)$ is the renormalized correlator of the disorder defined and studied in \cite{LeDoussalMiddletonWiese2008,LeDoussalWiese2008c} and measured in Ref. \cite{LeDoussalWieseMoulinetRolley2009}. It is $m$ dependent: at large $m$ it is equal to the bare disorder correlator, while as $m$ is decreased it develops a linear cusp at $u=0$, whose value, $\Delta'(0^+)$, is related to the second moment of the size distribution as displayed above in equation (\ref{70}). 

The scale $S_{m}$ is the large-scale cutoff for avalanche sizes, originating from the quadratic well
which suppresses the largest avalanches. It is an important scale as it allows to define universal
functions in the limit where it becomes large i.e.\ $S_m \gg S_{\mathrm{min}}$, where $S_{\mathrm{min}}$ is the typical size of the smallest avalanches. In the variable $s:=S/S_m$ the avalanche-size distribution
becomes universal. Universal means independent of small-scale
details, but not of the large-scale setting, e.g.\ it will depend on the precise form
of the elastic kernel. Indeed, one of the predictions of the FRG
theory is that if the exponent $\tau$ satisfies $2> \tau >1$ which
will turn out to be the
case here, then the distribution of avalanche sizes for $S \gg S_{\mathrm{min}}$ takes the form as $m \to 0$, i.e.\ $S_m \gg S_{\mathrm{min}}$,
\begin{equation}\label{a2}
P (S) \rmd S := \frac{\left< S \right>}{S_m} p \left(\frac{S}{S_{m}}\right)  \frac{\rmd S}{S_m}\ .
\end{equation}
The function $p(s)$ is universal in the above sense. For a fixed elastic kernel it depends only on the space dimension $d$. Note that the normalized
probability $P(S)$ depends on the cut-off $S_{\mathrm{min}}$ via the
first moment $\left< S \right>$ which cannot be predicted by the theory, hence is an input from experiment. It is important to stress that while the function
$p(s)$ is universal and convenient for data analysis, {\it it is not a probability distribution} and
is not normalized to unity. Rather, it satisfies from its definition (\ref{a2}) and using (\ref{70}) the two normalization conditions
\begin{eqnarray}\label{7}
\left< s \right>_p&=& \int \rmd s\, s~ p (s) =1\ ,\\
\left< s^{2} \right>_p &=&  \int \rmd s\, s^{2} p (s) =2 \ . \label{8} 
\end{eqnarray}
Here and below we use the notation $\left< s \right>_p$ to denote an integration over $p(s)$ and
distinguish it from a true expectation value over $P(S)$, denoted $\left<\ldots\right>$.
Note that $\tilde Z (\lambda)$ and $p (s)$ are related by the Laplace
transform 
\begin{equation}\label{k4}
\tilde  Z (\lambda) = \int_{0}^{\infty}\rmd s\, p (s)
\left[e^{\lambda s}-1 \right] \ .
\end{equation}
Note finally that the limit $m \to 0$ means that $L_c$ is large compared to the microscopic cutoff $a$ along the line, which may be of  nanometer scale, or for strong disorder the size of the defects. 

\subsubsection{Results from one-loop FRG}
\label{resultsFRG}

We now summarize the results obtained in \cite{LeDoussalWiese2008c} for
a general form of the elastic kernel $\epsilon_k$. One defines:
\begin{eqnarray}
 \tilde \epsilon_{k} := \epsilon_{k}/\epsilon_{0}  \  , \qquad \epsilon_{0}=m^2\ .
\end{eqnarray}
The general result of \cite{LeDoussalWiese2008c} is that $\tilde Z(\lambda)$ satisfies, in an expansion in powers of the renormalized disorder $\Delta(w)$ and up to terms of ${\cal O}(\Delta^2)$, 
\begin{eqnarray} \label{Ztform}
 \tilde Z&=& \lambda + \tilde Z^{2} + \alpha J(\tilde Z) \  , \qquad \alpha = - \epsilon \int_k \epsilon_k^{-2} \Delta''(0^+) \qquad \\
 J(z) &:=&  \frac{1}{\epsilon \int_k \tilde \epsilon_k^{-2}}  \int_{k} \bigg[ \frac{z^{2}}{(\tilde \epsilon_k-2 z)^{2}}
 + \frac{z}{\tilde \epsilon_k-2 z} -  \frac{z}{\tilde \epsilon_k} -  3 \frac{z^2}{\tilde \epsilon_k^2}\bigg] \label{88}\ .\nonumber \\
\qquad 
\end{eqnarray}
Note that the parameter 
\begin{equation}
\epsilon:=d_c-d\ ,
\end{equation} the distance to the upper critical dimension, introduced here for later purpose, cancels and is thus immaterial in this formula.  As such, this equation is formally exact for any $m$, any $\epsilon_k$ and any dimension, up to ${\cal O}(\Delta^2)$ terms.

Furthermore, consider now an elastic kernel such that:
\begin{equation}
 \epsilon_k \sim_{k \to \infty} K k^\beta
\end{equation}
with elasticity range $\beta \leq 2$ (i.e.\ long-ranged for $\beta<2$ and short-ranged for $\beta=2$) and $K$ the elastic constant. Assume that it takes the form
\begin{equation} \label{defscalee} 
 \tilde \epsilon_{k}  = e(K^{1/\beta} k/\mu) \  , \qquad \mu^\beta:=m^2\ ,
\end{equation}
where the elastic scaling function $e(p)$ is constructed as the unique dimensionless function of the dimensionless argument $p$ which satisfies $e(0)=1$, and $e(p) = p^\beta$ for $p \to  \infty$. Then, the result of  \cite{LeDoussalWiese2008c} is that the formula (\ref{Ztform}) can be controlled, in the limit $m \to 0$ as an expansion in powers of $\epsilon=d_c-d$, where $d_c=2 \beta$ is the upper critical dimension. For this one uses that
\begin{eqnarray}
\int_k \epsilon_k^{-2}  &=& \mu^{d- 2 \beta} K^{-d/\beta} \int_p e(p)^{-2}\nonumber \\
& =& \mu^{-\epsilon} K^{-2+ \frac{\epsilon}{\beta}}  \frac{C_\beta}{\epsilon} + {\cal O}(1) 
\end{eqnarray}
in the limit $\epsilon=d_c-d \to 0$, where $C_\beta= 2 (4 \pi)^{-d_c/2}/\rGamma(\beta)$ is {\it independent} of the shape of the function $e(p)$. Since the 1-loop FRG flow for $\Delta$ has the schematic form $- \mu \partial_\mu \Delta = - \mu \partial_\mu (\int_k \epsilon_k^{-2}) \times \Delta^2 = (\epsilon \int_k \epsilon_k^{-2})  \Delta^2$, it is convenient to introduce the
rescaled disorder,
\begin{equation} \label{defresc} 
 \tilde \Delta''(0^+) := (\epsilon \int_k \epsilon_k^{-2}) \Delta''(0^+) =  K^{-2} C_\beta \mu^{-\epsilon} \Delta''(0^+)\ ,
\end{equation}
where the second equality holds to leading order in $\epsilon$. This is precisely the parameter $\alpha$ defined
above, i.e.\ $\alpha = - \tilde \Delta''(0)$. The one-loop FRG flow for the rescaled disorder  admits a fixed point, from which one obtains, as $m \to 0$, that $\alpha$ flows to $\alpha = - \tilde \Delta''(0)= -
  \frac{1}{3} \epsilon (1- \zeta_1)$. Here $\zeta_1$ is the ${\cal O}(\epsilon)$ correction to the 
  roughness exponent of the elastic object. For the type of disorder relevant for the contact line (i.e.\ random-field disorder) one has $\zeta_1=1/3$, hence
\begin{equation}
\alpha=- \frac{2}{9} \epsilon\ .
\end{equation}
This value will be used  from now on  in (\ref{Ztform}}),  which is thus valid up to terms of order ${\cal O}(\epsilon^2)$. Using the above definitions one finds  that to lowest order in $\epsilon$, $J(z)$ depends only on the dimensionless scaling function $e(p)$, and not on the stiffness $K$:
\begin{equation}\label{k6n}
 J(z) = C_\beta^{-1} \int_p \bigg[ \frac{z^{2}}{(e(p)-2 z)^{2}}
 + \frac{z}{e(p)-2 z} -  \frac{z}{e(p)} -  3 \frac{z^2}{e(p)^2}\bigg] \ ,
\end{equation}
valid for SR elasticity as well as LR elasticity. Of course, since we work to lowest order in
$\epsilon$, the above integral should be computed in the critical dimension $d=d_c$. 
   
 Note that the mean-field singularity of this equation corresponds to $p=0$, $e(p=0)=1$, i.e.\ $z=1/2$.   
   
\subsection{From $\tilde Z (\lambda)$ to $p (s)$}\label{k7}

In Refs.\ \cite{LeDoussalMiddletonWiese2008,LeDoussalWiese2008c} we examined various
choices for the elastic kernel $\epsilon_k$, computed $\tilde Z(\lambda)$
for each choice, and  extracted the scaled avalanche distribution $p(s)$. 
Here we show that $p(s)$ can directly be written as a function of
$\epsilon_k$, or $e(p)$. 

\subsubsection{General formalism}\label{f4}
The definition (\ref{k4}) implies 
\begin{equation}\label{k8}
 \int_0^\infty ds\, s p(s) e^{\lambda s} = \tilde Z'(\lambda)\ .
\end{equation}
While $p(s)$ does not admit a Laplace transform,  the function $s p(s)$ does. Laplace inversion then yields
\begin{eqnarray}\label{f5}
 s p(s) &=& \frac{1}{2 i \pi} \int_{-i \infty}^{i \infty} d\lambda\, e^{- \lambda s}  \tilde Z'(\lambda) \nonumber \\
&=& \frac{1}{2 i \pi} \int_{-i \infty}^{i \infty} dZ\, e^{- s (Z - Z^2 - \alpha J(Z))} \ ,
\end{eqnarray}
up to terms of order $\alpha^{2}$. 
At this stage there is a heuristic step, to go from the $\lambda$ to
the $Z$ contour. We assume that the contour $Z= i x$ with $x\in
\mathbb{R}$ is the correct one, as it is for the mean-field case ($\alpha =0$)
around which we perturb. We will check this result  on known cases below. 
Then one has
\begin{eqnarray}\label{f6}
 s p(s) &=&  \frac{1}{2  \pi} \int_{-\infty}^{\infty} dx\, e^{- i s x - s x^2 + s \alpha J(i x)}  \\
&=& s p_{\mathrm{MF}}(s) +   \frac{\alpha s}{2  \pi} \int_{-\infty}^{\infty} dx \,J(i x) e^{- i s x - s x^2} \nn 
\end{eqnarray}
to lowest order in $\alpha$. We have introduced the scaled  mean-field avalanche-size distribution, i.e.\ (\ref{f5}) at $\alpha=0$, 
\begin{eqnarray}\label{f6}
p_{\mathrm{MF}}(s) = \frac{1}{2 \sqrt{\pi}} s^{-3/2} e^{-s/4} \ .
\end{eqnarray}
For the elastic manifold, it holds for $d \geq d_c$. We want to compute the correction to $s p(s)$ of order $\alpha$, i.e.\ of order $\epsilon=d_c-d$:
\begin{align}
& \frac{s}{2  \pi} \int_{-\infty}^{\infty} dx\, J(i x) e^{- i s x - s x^2} \nn \\
& = \frac{1}{\epsilon \int_k \tilde \epsilon_k^{-2}}   \frac{s}{2  \pi}  \int_0^\infty dt  \int_k  \int_{-\infty}^{\infty} dx\, \big[ t e^{- t (\tilde \epsilon_k - 2 i x)} \partial_y^2 
\nonumber \\
&+ e^{- t (\tilde \epsilon_k - 2 i x)} \partial_y - e^{- t \tilde \epsilon_k } \partial_y - 3 t e^{- t \tilde \epsilon_k } \partial_y^2 \big] 
 e^{- i s x - s x^2 + i y x} |_{y=0} \nn \\
 & = \frac{1}{\epsilon \int_k \tilde \epsilon_k^{-2}}   \frac{s}{2  \pi}  \int_0^\infty dt  \int_k  \int_{-\infty}^{\infty} dx\, 
 x e^{ - \tilde \epsilon_k t - i s x - s x^2} \nonumber \\
&\hphantom{\frac{1}{\epsilon \int_k \tilde \epsilon_k^{-2}}   \frac{s}{2  \pi}  \int_0^\infty dt  \int_k  \int_{-\infty}^{\infty} dx\,}\times \left[ e^{2 i t x} ( i - t x) + 3 t x - i\right] \ .\label{last}
\end{align}
The integrand behaves (before integration over $x$ and $k$) as $t^2$ at small $t$ as a result of the counterterms. It is useful to introduce the {\it elastic generating function}, 
\begin{equation}\label{Cdef}
{\cal C} (t):=\frac{1}{\epsilon \int_k \tilde \epsilon_k^{-2}} \int_k 
 e^{- t \tilde \epsilon_k} = \frac{1}{C_\beta} \int_p 
 e^{- t e(p)}\ ,
\end{equation}
in terms of which we get, integrating (\ref{last}) over $x$:
\begin{eqnarray}\label{87}
 \frac{p(s)}{p_{\mathrm{MF}}(s)} &=& 1 + \alpha \frac{1}{4 s} \int_0^\infty dt ~ {\cal C}(t) X (t,s)\\
X (t,s) &=&  e^{t  - t^2/s} \Big[ 4 t^3 + s^2 (2+t)- 2 s t (3 + 2 t) \Big] \nonumber \\
&&- s \Big[ 2 s + 3 t (s-2) \Big]
\ .
\end{eqnarray}
This is our final and most general formula, valid for any elasticity $\epsilon_k$ and to first order in $\alpha$, i.e.\ in $\epsilon$. By performing the integral over $s$, one checks that it  satisfies automatically the two normalization conditions for $p(s)$ (to order $\alpha$). 

\subsubsection{Special cases}\label{f7}

We now check that this formula recovers previous known results. 

\paragraph{Standard local elasticity:}

\medskip

For $\epsilon_k = K k^2 + m^2$ one has $\beta=2$ and $e(p)=p^2+1$. Using that $\int_k e^{- t k^2} =(4 \pi)^{-d/2} t^{-d/2}$ and computing the momentum integral in $d=d_c=4$, one finds
\begin{equation}\label{C1}
{\cal C}(t) = \frac1{2t^2} e^{-t}
\ .
\end{equation}
Inserting into (\ref{87}) and performing the $t$ integral yields
\begin{align} \label{sr}
& \frac{p(s)}{p_{\mathrm{MF}}(s)} = 1 +  \frac{ \alpha}{16} \left[(\ln s + \gamma_E ) (s-6)+4 s-8 \sqrt{\pi }
   \sqrt{s}+4\right],
\end{align}
which recovers the result (169) of  \cite{LeDoussalWiese2008c} (to lowest order in $\alpha$).

\paragraph{Elasticity of flat contact line and generalization:} 

From (\ref{f1}) the elasticity of a flat contact line $\varphi+\theta=\pi/2$ is $\epsilon_k = \gamma \sin^2 \theta \sqrt{k^2 + \kappa^2}$. This gives:
\begin{eqnarray}
&& \beta=1 \quad , \quad K=\gamma \sin^2 \theta \quad , \quad m^2=\mu=K/\kappa\qquad \\
&& e(p) = \sqrt{p^2 + 1} \ .
\end{eqnarray} 
It is instructive to slightly generalize this, and study
\begin{eqnarray} \label{kernela} 
 e(p) = \sqrt{p^{2}+ (1-a)^{2}}+a
\end{eqnarray} 
with $0<a<1$, which interpolates between the flat contact line for $a=0$ and
$e(p)=|p|+1$ for $a=1$, two cases studied in Ref. \cite{LeDoussalWiese2008c}. From the definition (\ref{Cdef}), computing the integral in $d=d_c=2$, 
we obtain
\begin{equation}\label{b3}
{\cal C} (t) =\frac{e^{-t}}{t^2}  \left[1+(1-a) t  \right]
\ .
\end{equation}
Inserting into (\ref{87}) and performing the $t$ integral yields
\begin{eqnarray}\label{f8}
 \frac{p(s)}{p_{\mathrm{MF}}(s)} = 1 +  \frac{\alpha}{8} \!&\Big\{&\!16 - 12 a - 6 \gamma_ {\mathrm{E}} + (1-a) \sqrt{\pi} s^{3/2} \nonumber \\
 &&\! + \left[(3-2 a)s- 6\right] \ln s  - 4 (3-a) \sqrt{\pi} \sqrt{s}   \nonumber \\
&&\! + s (2 a (5-\gamma_{\mathrm{E}}) + 3 (\gamma_ {\mathrm{E}}-2) \Big\}
\ .~~
\end{eqnarray}
For $a=0$, this is the same as (E14) of \cite{LeDoussalWiese2008c}, and for $a=1$
has the same form as (\ref{sr}) upon a rescaling of $\alpha$ by a factor of $2$, as noted in Ref. \cite{LeDoussalWiese2008c}.

To obtain nicer forms for $p(s)$, and more convenient for extrapolations in physical dimension,
one needs to re-exponentiate these direct $\epsilon$
expansion results, as done in Ref.\ \cite{LeDoussalWiese2008c}. We will not attempt
to do this here for the general case.

\subsection{More on the generating function $\tilde  Z (\lambda )$}\label{f9}

$\tilde Z(\lambda)$ is easier to  compare with numerical and experimental data than the disorder distribution $p(s)$, since even if their statistics is mediocre averages over all data are taken, thus the statistical fluctuations are less pronounced. Therefore, we  
come back to the general formula and some examples. 

\subsubsection{Generating function in the general case}

Since we work to the first order in $\alpha$, i.e.\ in $\epsilon=d_c-d$ 
we can write:
\begin{equation}\label{b1}
\tilde Z (\lambda) =\frac{1}{2} \left(1-\sqrt{1-4 \lambda
}\right) + \alpha \frac{J \left (\frac{1}{2} [1-\sqrt{1-4 \lambda
}] \right)}{\sqrt{1-4\lambda }}
\Big|_{z=Z_{\mathrm{MF} (\lambda)}}  
\ .
\end{equation}
up to higher order terms, 
hence we only need to compute the function $J(z)$ in eq.\ (\ref{88}) at the 
point of the mean field solution $z=Z_{\mathrm{MF}}(\lambda)= \frac{1}{2} \left(1-\sqrt{1-4 \lambda
}\right)$.

The integral $J(z)$ can be rewritten using the elastic generating function 
${\cal C} (t)$ defined in (\ref{Cdef}) for an arbitrary elastic kernel $\epsilon_k$ as
\begin{equation}\label{a26}
J (z)=\int_{0}^{\infty }dt\, {\cal C} (t)\, \left[\rme^{2 t z}\left (
tz^{2}+ z \right)- z -3 t z^{2} \right]\ .
\end{equation}

\subsubsection{Standard elasticity}

Standard elasticity has the form $e(p)=p^2+1$ and using (\ref{C1}) one finds:
\begin{eqnarray} \label{ref3} 
J(z) &=& \frac{1}{2} z \left[2 z+(1-3 z) \log (1-2 z) \right]\ .
\end{eqnarray}
This result is in agreement with eq.~(150) of \cite{LeDoussalWiese2008c}.
The value of $\alpha$ to be used in (\ref{b1}) for extrapolation to $d=1$ (i.e.\ $\epsilon=3$)
 is $\alpha=-2/3$. 
\subsubsection{Flat contact line and generalization }

From the form $e(p) = \sqrt{p^2+(1-a)^2} + a$ and using (\ref{b3}) one finds
\begin{equation}\label{106}
J (z) = z \left[\frac{2 z (a+z-3 a z)}{1-2 z}+(a-3 z) \log (1-2 z)\right]\ .
\end{equation}
This agrees with (E10) of \cite{LeDoussalWiese2008c} for $a=0$ and with (\ref{ref3}) up to a global
factor of $2$, as predicted in \cite{LeDoussalWiese2008c}. The value of $\alpha$ to be used in (\ref{b1}) for extrapolation to $d=1$  (i.e.\ $\epsilon=1$) is $\alpha=-2/9$.

\subsubsection{Contact line for arbitrary angles $\theta$ and $\varphi$}

Having tested our general formula on the known cases we can now apply them to
the case of the elastic kernel (\ref{final}). Let us specify the relevant parameters and functions. For the contact line $\beta=1$, $d_c=2$, and\begin{eqnarray}
 m^2&=&\mu= \epsilon_{q=0} 
= \frac{\gamma \kappa}{t} \sin \theta  \cos \varphi \\
 K &=&  \gamma \sin^2 \theta
 \\
 \kappa&=& \frac{1}{L_c} = \sqrt{\frac{\rho g}{\gamma}} \quad , \quad  t= \sqrt{\frac{\sin(\theta +\varphi )+1}{2}}  \\
 \tilde \epsilon_q &=& e(p) = \tilde e(r) = 1 +   \frac{\left(r^2-1\right)\left [t
 (r+t)+1\right]\sin \theta}{  \left(r^2+3 r t+3 t^2-1\right) \cos \varphi} \qquad \ \ \\
 r &=& \sqrt{1+ \frac{q^2}{\kappa^2}}  = \sqrt{1+ \frac{\cos^2 \varphi}{t^2 \sin^2 \theta} ~ p^2} \label{correct}\ , \quad p=\frac K \mu q\ .
\end{eqnarray}
We have defined for convenience a new function $\tilde e(r)$ since $r$ is the implicit variable in the
final formula. Note that for the flat interface $\theta+\varphi=\pi/2$, one recovers $\tilde e(r)=r$. 

The generating function $\tilde Z(\lambda)$ for the contact line can thus be computed from (\ref{b1})
using (\ref{k6n}) and performing the integral in dimension $d_c=2$. Upon a change of variable we get:
\begin{eqnarray}\label{k6}
  J(z)  &=& \frac{t^2 \sin^2 \theta}{\cos^2 \varphi}\\
&&\times   \int\limits_1^{\infty}\rmd r\, r \bigg[ \frac{z^{2}}{[\tilde
e (r){ -}2 z]^{2}}
 {+} \frac{z}{\tilde e (r) {-}2 z}  {-}  \frac{z}{\tilde e (r)} {-}  3 \frac{z^2}{\tilde e (r)^2}\bigg] \ . \nonumber
\end{eqnarray}
Performing this integral analytically, or even the one involved in computing ${\cal C}(t)$, is rather awkward, and we have preferred to use numerical integration. 

\begin{figure}
\fig{1}{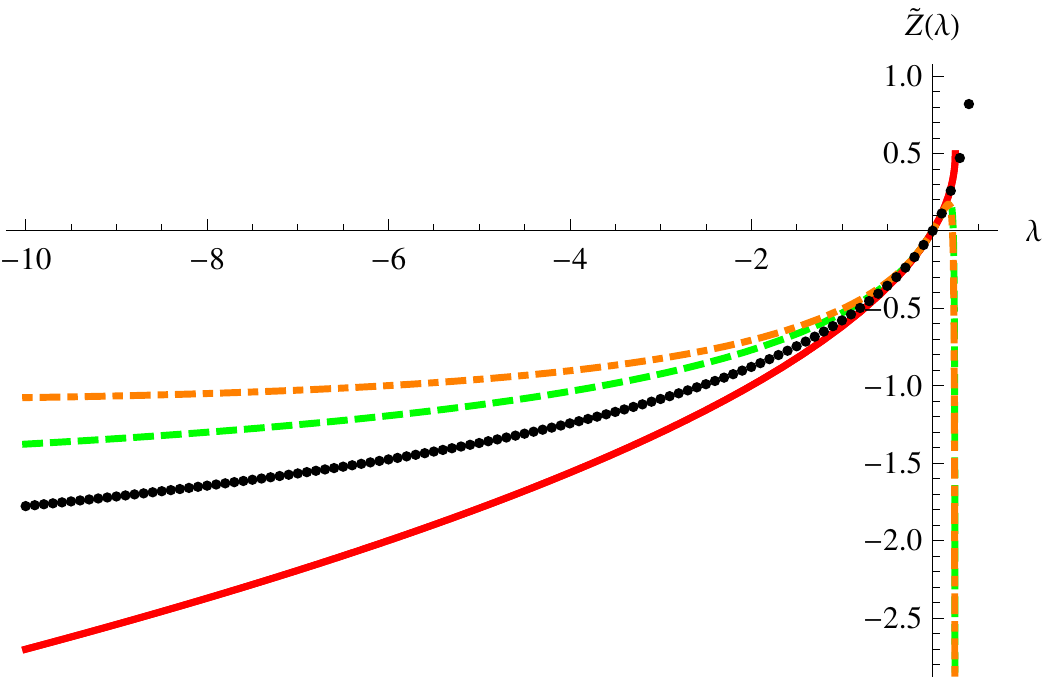}
\caption{The generating function $\tilde Z (\lambda)$ of avalanche sizes as defined in (\ref{k4}). The curves are from bottom to top: Red/solid: mean-field result. Black dots: The experimental data of \cite{LeDoussalWieseMoulinetRolley2009}. 
Dashed/green: Eq.~(\ref{k6}) integrated numerically with the correct elasticity kernel (\ref{correct}) for $\theta
=40^{\circ}$, $\varphi =0^{\circ}$. Dot-Dashed/orange: The same with the elastic kernel for $\theta
=90^{\circ}$, $\varphi =0^{\circ}$, $\tilde \epsilon_q=\sqrt{1+q^2/\kappa^2}$. 
%
}
\label{f:tZlambdaVRAI}
\end{figure}

To relate to the experiment of \cite{LeDoussalWieseMoulinetRolley2009}, we have computed $J(z)$ numerically for $\theta
=40^{\circ}$, $\varphi =0^{\circ}$. We give some explicit values: $J
(-1/2)=-0.493277$, $J (-1) =-3.10539$, $J (-3/2)=-8.79405$, $J
(-2)=-18.1329$. From this we have computed and plotted $\tilde Z (\lambda)$ in 
figure \ref{f:tZlambdaVRAI}, 
using the extrapolation to $\epsilon=1$ in the one-loop result, i.e.\ using (\ref{b1}) with $\alpha
=-2/9$. This is compared to the result of  the scaled kernel $e(p)=\sqrt{p^2 + 1}$. The experimental data are also plotted on figure \ref{f:tZlambdaVRAI}. Note that using the correct elasticity 
allows to get closer to the experimental data than using $e(p)=\sqrt{p^2+1}$, which is the kernel
for the flat interface, and the only one previously available in the literature.

\subsection{Local statistics of avalanches}

\subsubsection{Basic definitions}\label{f10}
It is also possible to study in experiments the statistics of the avalanches occurring within a given portion of the 
elastic object, e.g.\ of the contact line. One thus defines the size of a local avalanche, weighted by a characteristic function $\phi(x)$, e.g.\ a $\delta$-distribution localized at $x=0$, as \footnote{Note that the variable $\phi(x)$ entering in $S^\phi$ and $a^\phi$ bears no relation to the plate angle $\varphi$ introduced previously.}:
\begin{eqnarray}\label{k3+}
S^\phi&:=& \int_{0}^{L}\rmd^{d} x\,  \phi(x) \left[u_{w}^{+} (x)-u_{w}^{-} (x)\right] \nonumber \\
&=& \int_{0}^{L}\rmd^{d} x\,  \phi(x) \left[h_{w}^{+} (x)-h_{w}^{-} (x)\right]\ .
\end{eqnarray}
In \cite{LeDoussalWiese2008c}, we have shown how the characteristic
function 
\begin{equation}\label{f11}
Z^{\phi} (\lambda) := \frac{1}{\left< S^\phi \right> } \left<
\rme^{\lambda S^{\phi }} -1 \right> 
\end{equation}
as well as the distribution of sizes can be computed within mean-field theory, valid at and above the
upper critical dimension $d_c$. Since it already required some quite involved instanton calculus, we did not attempt to perform the $\epsilon$ expansion. Here we will perform the $\epsilon$ expansion, to one loop ${\cal O}(\epsilon)$, not on the full distribution, but on a simpler quantity, the second moment. More precisely, we will compute the universal amplitude ratio between the second global and local moments, quantities easily accessible in experiments and numerics. In fact, it was already measured in \cite{RossoLeDoussalWiese2009a} in the case of standard local elasticity and the present analytical result was quoted. Here we provide the details of the calculation and predict this ratio for the contact line which has not yet been measured.

\subsubsection{Universal amplitude ratios between global and local avalanches: general
formulation} \label{yy3}

Analogous to (\ref{70}), we can define the scale of local avalanches
as 
\begin{equation}\label{f12}
S^\phi_m := \frac{\left<(S^\phi)^2\right>}{2 \left<S^\phi\right>}
\ .
\end{equation}
From Eq. (F12) and (196) in \cite{LeDoussalWiese2008c} one has the general exact relation:
\begin{equation}\label{f13}
S^\phi_m
= -  \frac{1}{\int_x \phi(x)} \partial_w \int_{x y z z'} \phi(y) \phi(x) g_{xz} g_{yz'} \Delta_{zz'}[w] \Big|_{w=0^+} \nonumber 
\ ,
\end{equation}
where $g_{xy}=\int_q e^{i q(x-y)} \epsilon_q^{-1}$ and $\Delta_{zz'}[w]  = - R''_{zz'}[w]$ is the functional derivative taken at a uniform $w_t=w$ of the functional $R[w]$ (see e.g.\ \cite{LeDoussal2006b,LeDoussalWiese2008c} for its definition), the limit $w=0^+$ being taken at the end. For $\phi(x)=1$, using $\int_{zz'} \Delta_{zz'}[w]=L^d \Delta'(w)$ one recovers
\begin{equation}\label{f14}
 S_m = \frac{\left<S^2\right>}{2 \left<S\right>} = -  m^{-4} \Delta'(0^+)  \ . 
\end{equation}
Here we use the notations of section (\ref{resultsFRG}) so that $m^2 =\epsilon_{k=0}$.

We can now use the results of the $\epsilon$-expansion \cite{ChauveLeDoussalWiese2000a,LeDoussalWieseChauve2002,LeDoussalWieseChauve2003}, or its first-principle 
derivation in formula (463) of \cite{LeDoussal2008}:
\begin{eqnarray}
 \Delta_{zz'}[w] = \delta_{zz'} \Delta(w) - (g_{zz'}^2 - \delta_{zz'} \int_t g_{t}^2) \Delta'(w)^2 + {\cal O}(\epsilon^3) \nonumber 
\ .
\end{eqnarray}
Hence one obtains the ratio
\begin{eqnarray}
 a_\phi &:=& \frac{S^\phi_m}{S_m} \\
&=& \frac{m^4}{\int_x \phi(x)} \int_{x y z} \phi(y) \phi(x) g_{xz} g_{yz} \nonumber \\
&& -  2 m^4 \Delta''(0^+) \frac{1}{\int_x \phi(x)} \nonumber\\
&& \times \int_{x y z z'} \phi(y) \phi(x) g_{xz} g_{xz'}  (g_{zz'}^2 - \delta_{zz'} \int_t g_{t}^2) + {\cal O}(\epsilon^2)\ , \nonumber 
\end{eqnarray}
a formula valid for any function $\phi(x)$. 

We know study avalanches which occur on a given subspace of co-dimension $d'$, defined by $x_\perp=0$, i.e.\ of dimension $d_\phi=d-d'$, of the $d$-dimensional manifold, i.e.\ we choose
\begin{eqnarray} \label{defphi} 
\phi(x)=\eta  \delta^{d'}(x_\perp)\ .
\end{eqnarray}
$\eta$ will be specified later. 
The above formula leads to
\begin{eqnarray}\label{f15}
 a_\phi&=&  m^4 \eta  I_{2} (d') \nn\\&&+ 2 m^4 \eta 
\left [I_{2} (d) I_{2} (d') -
 I_{A} (d,d') \right] \Delta'' (0) + \dots\nonumber \\
 &=& m^4 \eta  I_{2} (d')\left[1+  \frac{2 \tilde \Delta'' (0)}{\epsilon}  \left ( 1 -
\frac{I_{A} (d,d')}{ I_{2} (d) I_{2} (d')}  \right)
 \right] +\dotsb  \nonumber \\ 
\end{eqnarray}
using the definition (\ref{defresc}) of the rescaled disorder, 
$\tilde \Delta''(0)=\Delta''(0) \epsilon I_2(d)$, and defining the integrals
\begin{eqnarray}\label{f16}
I_{2} (d) &:=& \parbox{1.5cm}{\includegraphics[width=1.5cm]{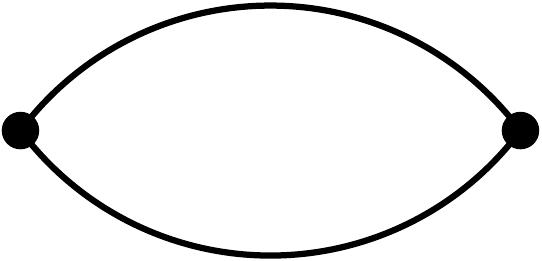}} = \int\frac{\rmd^{d}q}{(2\pi)^{d}} \frac{1}{\epsilon_{q}^{2}}\\
I_{A} (d,d') &:=& \parbox{1.cm}{\includegraphics[width=1.cm]{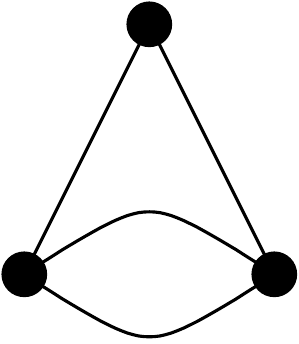}}\nonumber \\
&=&\int \frac{\rmd^{d'} k}{( 2\pi)^{d'}}
\frac{\rmd^{d}q}{(2\pi)^{d}} \frac{1}{\epsilon_{k}^{2}}
\frac{1}{ \epsilon_{q}} \frac{1}{\epsilon_{k+q}}
\ .
\end{eqnarray}
Note that in the last integral the momentum in the upper bubble runs
over a space of dimension $d'<d$.

To perform the $\epsilon$ expansion we now use the scaling form (\ref{defscalee}) of the
elastic kernel. This yields to zeroth and first order in $\epsilon=d_c-d$:
\begin{eqnarray}\label{f15}
 a_\phi&=& \eta  \mu^{d'} K^{-d'/\beta} \tilde I_{2} (d')\left[1+  \frac{2 \tilde \Delta'' (0)}{\epsilon}  \left ( 1 -
\frac{\tilde I_{A} (d,d')}{ \tilde I_{2} (d) \tilde I_{2} (d')}  \right)
 \right] \nonumber \\
 && +\dotsb  
\end{eqnarray}
in terms of the dimensionless integrals
\begin{eqnarray}\label{f16}
\tilde I_{2} (d) &:=& \parbox{1.5cm}{\includegraphics[width=1.5cm]{./figures/dia1}} = \int\frac{\rmd^{d}q}{(2\pi)^{d}} \frac{1}{e(q)^{2}}\\
\tilde I_{A} (d,d') &:=& \parbox{1.cm}{\includegraphics[width=1.cm]{./figures/LH}}\nonumber \\
&=&\int \frac{\rmd^{d'} k}{( 2\pi)^{d'}}
\frac{\rmd^{d}q}{(2\pi)^{d}} \frac{1}{e(k)^{2}}
\frac{1}{e(q)} \frac{1}{e(k+q)}
\ .\qquad
\end{eqnarray}
To obtain a meaningful amplitude we now make the choice
\begin{eqnarray}
\eta  = \left(\frac{K^{1/\beta}}\mu\right)^{\!d'}\ .
\end{eqnarray}
Note that comparing with experiments then requires a precise knowledge of the internal length scale $K^{1/\beta}/\mu$
defined by the (renormalized) mass $m^2=\mu^\beta$ 
and elastic constant $K=\lim_{q \gg \kappa} \epsilon_q/q^\beta$. 
Alternatively, it may be used as a method to measure this length.

\subsubsection{Standard local elasticity}

We know specify to standard elasticity $\epsilon_k = K k^2 + m^2$, i.e.\ $\beta=2$, $d_c=4$ and $e(p)=p^2+1$. We study $d'=1$.

We need the integrals:
\begin{eqnarray}\label{f17}
 \tilde I_2(d) &=&  \rGamma \! \left(2-\frac{d}{2}\right) (4 \pi)^{-d/2} \\
 \tilde I_2(d'=1) &=& \int \frac{d q}{2 \pi} \frac{1}{(q^2 + 1)^2} = \frac{1}{4}  \\
 \label{f179}
\frac{\tilde I_{A} (d,d')}{ \tilde I_{2} (d) \tilde I_{2} (d')} &-& 1 = \frac{I}{ \tilde I_{2} (d) \tilde I_{2} (d')}
\end{eqnarray}
This involves the integral
\begin{eqnarray}
  I &=&  \int \frac{d k}{2 \pi} \frac{1}{(k^2 + 1)^2} \int \frac{d q}{2 \pi} \frac{d^{d-1} p}{(2 \pi)^d}  \\
&& \times \left[ \frac{1}{[(k+q)^2 + p^2 + 1](q^2 + p^2 + 1)} - \frac{1}{(q^2 +p^2 + 1)^2} \right] \!\nonumber
\end{eqnarray}
The calculation of this integral is performed in Appendix \ref{a:integral}. To the order $\epsilon$ 
we are working, we only need to compute it in $d=d_c=4$. We find
\begin{equation}\label{f19}
 I = (4 \pi)^{- \frac{5}{2}}\, \rGamma\!\left(\frac{3}{2}\right) \left[ 4 + \frac{2}{9} (7 \sqrt{3}-18) \pi \right]
\ .
\end{equation}
Putting all these results together we obtain
\begin{eqnarray}\label{f20}
 a_\phi&=& \frac{1}{4} +  \alpha  \left ( 1 -\pi +  \frac{7\pi }{6 \sqrt{3}} \right) 
 +{\cal O}(\epsilon^2) \nonumber \\
&=& 0.25 - 0.0254934 \alpha  +{\cal O}(\epsilon^2)  
\end{eqnarray}
with $\alpha=-\frac{\epsilon}{3} (1-\zeta_1)$. This is the expansion for a  manifold with co-dimension one in $\epsilon=4-d$. 

For the 1-dimensional string, simulated in
\cite{RossoLeDoussalWiese2009a}, 
we need to extrapolate to $\epsilon=3$ which yields
$\alpha= - 2/3$. The two possible Pad\'e
expansions are 
\begin{eqnarray}
\frac1{a_\phi}&=& \frac1{\frac{1}{4} + 0.0169956} = 3.74538 \ ,\\
\frac1{a_\phi}&=& 4 - 16  \alpha  \left[ 1 +  \left(\frac{7}{6 \sqrt{3}} -1\right) \pi \right]  = 3.72807\ .\qquad 
\ .
\end{eqnarray}
Hence an estimate of the ratio is
\begin{eqnarray} \label{epsrat}
 \frac1{a_\phi}= 3.74 \pm 0.01\ .
\end{eqnarray}

Another expansion is the  expansion in fixed dimension in powers of $\tilde \Delta$. It is likely to be
less precise than the $\epsilon$-expansion, but being simpler in spirit we indicate here its prediction.
One evaluates directly the integrals in $d=d'=1$. Using the
results from appendix \ref{a:integral}, this yields 
\begin{eqnarray}\label{f21}
a_{\phi} &=&  \frac{1}{4}\left[1+ \frac{2}{9} \frac{\tilde \Delta''(0^+)}{\epsilon} +\dots \right]\ ,
\end{eqnarray}
and inserting the one-loop value $\tilde \Delta''(0^+)=-\alpha=\frac{2}{9} \epsilon$ we find
the estimate $1/a_\phi \approx 3.81$, slightly larger than (\ref{epsrat}).

\subsubsection{Contact-line elasticity}

For the contact-line elasticity, the $\epsilon=2-d$ expansion can be done as indicated above. As it requires 
performing two-dimensional integrals we will only give here the fixed-dimension estimate for $d=d'=1$, which requires only a one-dimensional integral on each momentum. We further perform them numerically using the form (\ref{correct}).
We give two examples:

(i) flat interface $\varphi+\theta=\pi/2$.

Then  $e(p)= \sqrt{p^2+1}$ and we obtain
\begin{eqnarray}\label{f23}
&& \tilde I_{2} (1) = \frac{1}{2} \\
&& 2\left[1-\frac{\tilde I_{A} (1,1)}{\tilde
I_2 (1)\tilde I_{2} (1)} \right]\approx  0.312811 \ldots
\end{eqnarray}
This yields
\begin{eqnarray}\label{f23}
 a_\phi =  \frac{1}{2}  \left(1 + 0.312811 \frac{2}{9} + \ldots \right) 
\end{eqnarray}
and $1/a_\phi \approx 1.87$.

\medskip

(ii) experiments of Ref. \cite{LeDoussalWieseMoulinetRolley2009}, i.e.\ $\varphi=0^\circ$, $\theta=40^\circ$:

We find
\begin{eqnarray}\label{f23}
&& \tilde I_{2} (1) \approx 0.41867 \\
&& 2\left[1-\frac{\tilde I_{A} (1,1)}{\tilde
I_2 (1)\tilde I_{2} (1)} \right]\approx 0.333101 \dotsb 
\end{eqnarray}
This yields
\begin{eqnarray}\label{f23}
 a_\phi =  0.41867 \left(1 + 0.333101 \frac{2}{9} + \ldots \right) 
\end{eqnarray}
and $1/a_\phi \approx 2.22$.

It is interesting to note that the ``elastic length'' $\eta=K/\mu$ which enters
the definition (\ref{k3+}), (\ref{defphi}) of $S^\phi$ and $a_\phi$ reads:
\begin{equation}
\eta = \frac K \mu = L_c \, \frac{\sin \theta }{\cos \varphi}\sqrt{\frac{\sin(\theta +\varphi )+1}{2}}
\end{equation}
i.e., it only involves the capillary length and geometric prefactors, thus is well-known 
experimentally. In addition, one can  measure the dependence of $a_\phi$
on the plate angle $\varphi$.

\section*{Acknowledgments}\label{f26}
It is a pleasure to thank Etienne Rolley and Alberto Rosso for
stimulating discussions. 

\appendix

\begin{figure*}
\section{Standard elasticity: Calculation of an integral}
\label{a:integral}

The integral defined in the text can be written as
\begin{eqnarray}\nn
I &=& \int \frac{d k}{2 \pi}  \int \frac{d q}{2 \pi} \frac{d^{d-1} p}{(2 \pi)^d}
 \int_{t,t_1,t_2>0} t e^{- t (k^2+1)}\left[ e^{-t_1 ((k+q)^2 + p^2 + 1) + t_2 (q^2+p^2+1)} - 
 e^{- (t_1+t_2) (q^2+p^2+1)}\right] \\
&& =(4 \pi)^{-\frac{d+1}{2}} \int_{t,t_1,t_2>0} t (t_1+t_2)^{- \frac{d-1}{2}} \left[ (t t_1 + t t_2 + t_1 t_2)^{-1/2} 
- (t (t_1+ t_2))^{-1/2} \right] \rme^{-t-t_{1}-t_{2}}\ ,
\end{eqnarray}
using the general formula $\int_q e^{- q \cdot C \cdot q} = (4 \pi)^{-d/2} (\det C)^{-1/2}$. Performing the changes of 
variables $t_1 \to t t_1$ and $t_2 \to t t_2$ yields
\begin{equation}
I = (4 \pi)^{-\frac{d+1}{2}} \int_0^\infty \rmd t \int_0^\infty\rmd t_1 \int_0^\infty \rmd t_2 \,
e^{-t \left(t_1+t_2+1\right)} 
   t^{\frac{5}{2}-\frac{d}{2}} \left(t_1+t_2\right){}^{\frac{1}{2}-\frac{d}{2}}
   \left[\frac{1}{\sqrt{t_2 t_1+t_1+t_2}} - \frac{1}{\sqrt{t_1+t_2}}\right] 
\ .
\end{equation}
Integration over $t$ yields
\begin{equation}
I = (4 \pi)^{-\frac{d+1}{2}}  \int_0^\infty\rmd t_1 \int_0^\infty \rmd t_2 
 \left(t_1+t_2\right){}^{\frac{1}{2}-\frac{d}{2}} \left(t_1+t_2+1\right){}^{\frac{d-7}{2}} \rGamma \left(\frac{7}{2}-\frac{d}{2}\right)
   \left[\frac{1}{\sqrt{t_2 t_1+t_1+t_2}} - \frac{1}{\sqrt{t_1+t_2}}\right]
\ .
\end{equation}
Setting $t_{1}=st$, $t_{2}= (1-s)t$, one obtains
\begin{eqnarray}\label{f28}
I &=& (4 \pi)^{-\frac{d+1}{2}}  \int_0^\infty\rmd t \int_0^1 \rmd s\, 
    t^{1-\frac{d}{2}}
   (t+1)^{\frac{d-7}{2}} \rGamma
   \left(\frac{7}{2}-\frac{d}{2}\right) \left[\frac{1}{\sqrt{1+(1-s) s t}} - 1\right] \nn \\
   &=& (4 \pi)^{-\frac{d+1}{2}}  \int_0^\infty\rmd t\, 
    t^{1-\frac{d}{2}}
   (t+1)^{\frac{d-7}{2}} \rGamma
   \left(\frac{7}{2}-\frac{d}{2}\right) \left[\frac{2}{\sqrt{t}} ~ \mbox{arccot}\left(\frac{2}{\sqrt{t}}\right) - 1\right]\ .
\end{eqnarray} 
One can perform the integral directly in $d=1$, with the result
\begin{eqnarray}
I = - \frac{1}{144}\ .
\end{eqnarray} 
Thanks to the counterterm it also admits a limit for $d=4$:
\begin{eqnarray}
I =  \frac{1}{8 \pi^2} \left[ \frac{1}{2} + \frac{\pi}{36}(-18 + 7 \sqrt{3}) \right]\ .
\end{eqnarray} 
Finally 
the expression for any $d$ is 
 \begin{eqnarray}
I &=& \frac{  9 \sqrt{3} 2^d \rGamma \left(2-\frac{d}{2}\right) \,
   _3F_2\left(\frac{1}{2},1,2-\frac{d}{2};-\frac{1}{2},\frac{3}{2};\frac{1}{4}\right)+2 \sqrt{\pi } \left(9
   \sqrt{3} 2^d-4 3^{d/2} (d+3)\right) \rGamma
   \left(\frac{3}{2}-\frac{d}{2}\right)}{ 4^{d+1} \pi ^{d/2} 9 \sqrt{3}} - 2^{-2-d} \pi^{-d/2} \rGamma\left(2-\frac{d}{2}\right)\ . \qquad
   \end{eqnarray} 
  The combination $2\left[1-\frac{\tilde I_{A} (d,1)}{\tilde
I_2 (1)\tilde I_{2} (d)} \right]$ is plotted on Fig.~\ref{f:1mAsB}.
\end{figure*}
\begin{figure}[tbh]
\Fig{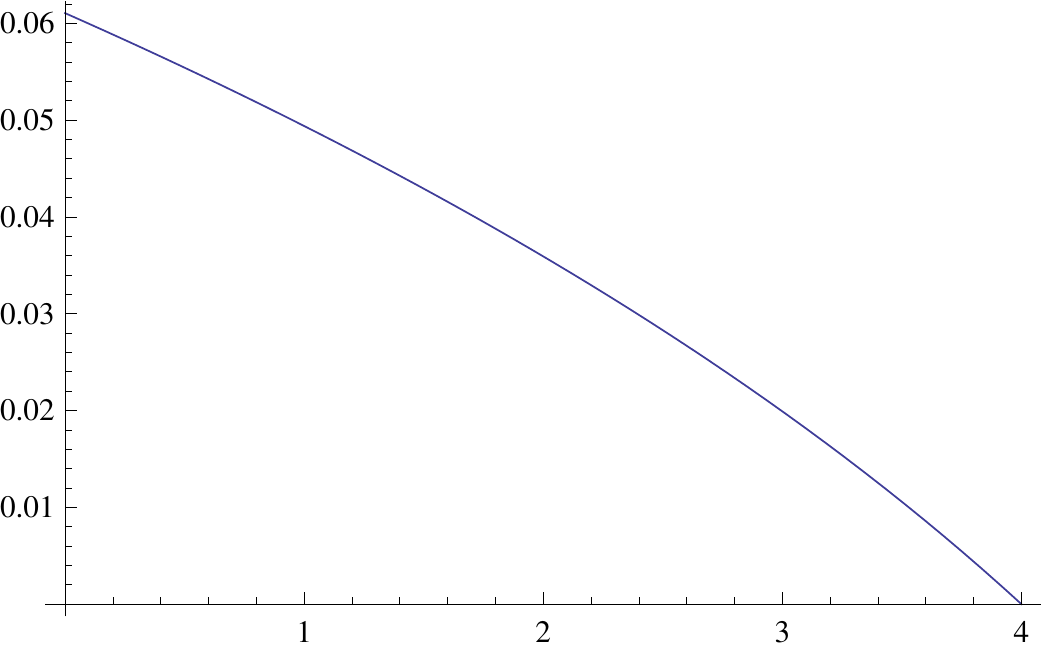}
\caption{The combination $2\left[1-\frac{\tilde I_{A} (d,1)}{\tilde
I_2 (1)\tilde I_{2} (d)} \right]$ as a function of $d$, for standard elasticity.}
\label{f:1mAsB}
\end{figure}

\newpage


%
\newpage
\tableofcontents
\end{document}